\documentclass{elsarticle}
\makeatletter
\def\ps@pprintTitle{%
 \let\@oddhead\@empty
 \let\@evenhead\@empty
 \def\@oddfoot{}%
 \let\@evenfoot\@oddfoot}
\makeatother
\usepackage{natbib}
\usepackage{geometry}
\usepackage{hyperref}
\usepackage{graphicx} 
\usepackage{amsfonts}
\usepackage{amssymb}
\usepackage{mathtools}
\usepackage{mathrsfs}
\usepackage{txfonts}
\usepackage{dsfont}
\usepackage{pifont}
\usepackage{subcaption}
\usepackage{booktabs}
\usepackage{xcolor}
\usepackage{tabularx}
\usepackage{multirow}
\usepackage{multicol}
\usepackage{cleveref}
\usepackage{float}
\usepackage{arydshln}
\usepackage{tikz,pgfplots}
\pgfplotsset{compat=1.18}  
\usepackage{bbm}
\usepackage{changes}
\definechangesauthor[name=TaoLi, color=orange]{TL}
\definechangesauthor[name=ZilinBian, color=red]{ZB}
\usepackage{cleveref}
\usepackage{bm}
\usepackage{tikz,pgfplots}
\pgfplotsset{compat=1.18}

\newcommand{\E}{\mathbb{E}}

\newcommand{\N}{\mathbb{N}}
\newcommand{\R}{\mathbb{R}}

\DeclareMathOperator*{\argmin}{arg\,min}

\Crefname{equation}{Eq.}{Eqs.}

\begin{document}
\begin{frontmatter}

\title{Digital Twin-based Driver Risk-Aware Intelligent Mobility Analytics for Urban Transportation Management}

\author[1]{Tao Li\fnref{contrib}}
\ead{tl2636@nyu.edu}
\author[2]{Zilin Bian\fnref{contrib}, \corref{poc}}
\ead{zb536@nyu.edu}
\author[1]{Haozhe Lei\fnref{contrib}}
\ead{hl4155@nyu.edu}
\author[2]{Fan Zuo\fnref{contrib}}
\ead{fan.zuo@nyu.edu}
\author[1]{Ya-Ting Yang\fnref{contrib}}
\ead{yy4348@nyu.edu}
\author[1]{Quanyan Zhu}
\ead{qz494@nyu.edu}
\author[3]{Zhenning Li}
\ead{zhenningli@um.edu.mo}
\author[4]{Zhibin Chen}
\ead{zc23@nyu.edu}
\author[2]{Kaan Ozbay}
\ead{kaan.ozbay@nyu.edu}

\fntext[contrib]{Authors contributed equally}
\cortext[poc]{Corresponding author}

\address[1]{Department of Electrical and Computer Engineering,  New York University}
\address[2]{Department of Civil and Urban Engineering, New York University}
\address[3]{State Key Laboratory of Internet of Things for Smart City, University of Macau}
\address[4]{Shanghai Key Laboratory of Urban Design and Urban Science, NYU Shanghai}

\begin{abstract} 

\noindent Traditional mobility management strategies emphasize macro-level mobility oversight from traffic-sensing infrastructures, often overlooking safety risks that directly affect road users. To address this, we propose a Digital Twin-based Driver Risk-Aware Intelligent Mobility Analytics (DT-DIMA) system. The DT-DIMA system introduces four novel services that extract real-time traffic information from pan-tilt-cameras (PTCs), synchronize this data into a digital twin to accurately replicate the physical world, and predict network-wide mobility and safety risks in real time. The system's innovation lies in its integration of advanced machine learning, simulation, and online control modules, including Spatial-Temporal Traffic Estimation (STTE) for real-time traffic data fusion and prediction, Mesoscopic Traffic Safety Simulation (MTSS) for real-time safety risk prediction, and Risk-Constrained Correlated Online Learning (RiCCOL) for real-time PTC control. Tested and evaluated under normal traffic conditions and incidental situations (e.g., unexpected accidents, pre-planned work zones) in a simulated testbed in Brooklyn, New York, DT-DIMA demonstrated mean absolute percentage errors (MAPEs) ranging from $8.40\%$ to $15.11\%$ in estimating network-level traffic volume and MAPEs from $0.85\%$ to $12.97\%$ in network-level safety risk prediction. In addition, the highly accurate safety risk prediction enables PTCs to preemptively monitor road segments with high driving risks before incidents take place. Such proactive PTC surveillance creates around a 5-minute lead time in capturing traffic incidents. The DT-DIMA system enables transportation managers to understand mobility not only in terms of traffic patterns but also driver-experienced safety risks, allowing for proactive resource allocation in response to various traffic situations. To the authors' best knowledge, DT-DIMA is the first urban mobility management system that considers both mobility and safety risks based on digital twin architecture.

\end{abstract}
\begin{keyword}
    Digital twin, traffic management and monitoring, real-time traffic simulation, online learning control, spatial-temporal forecasting.
\end{keyword}
\end{frontmatter}
\section{Introduction}

\begin{figure}[t!]
    \centering
    \includegraphics[width=0.97\columnwidth]{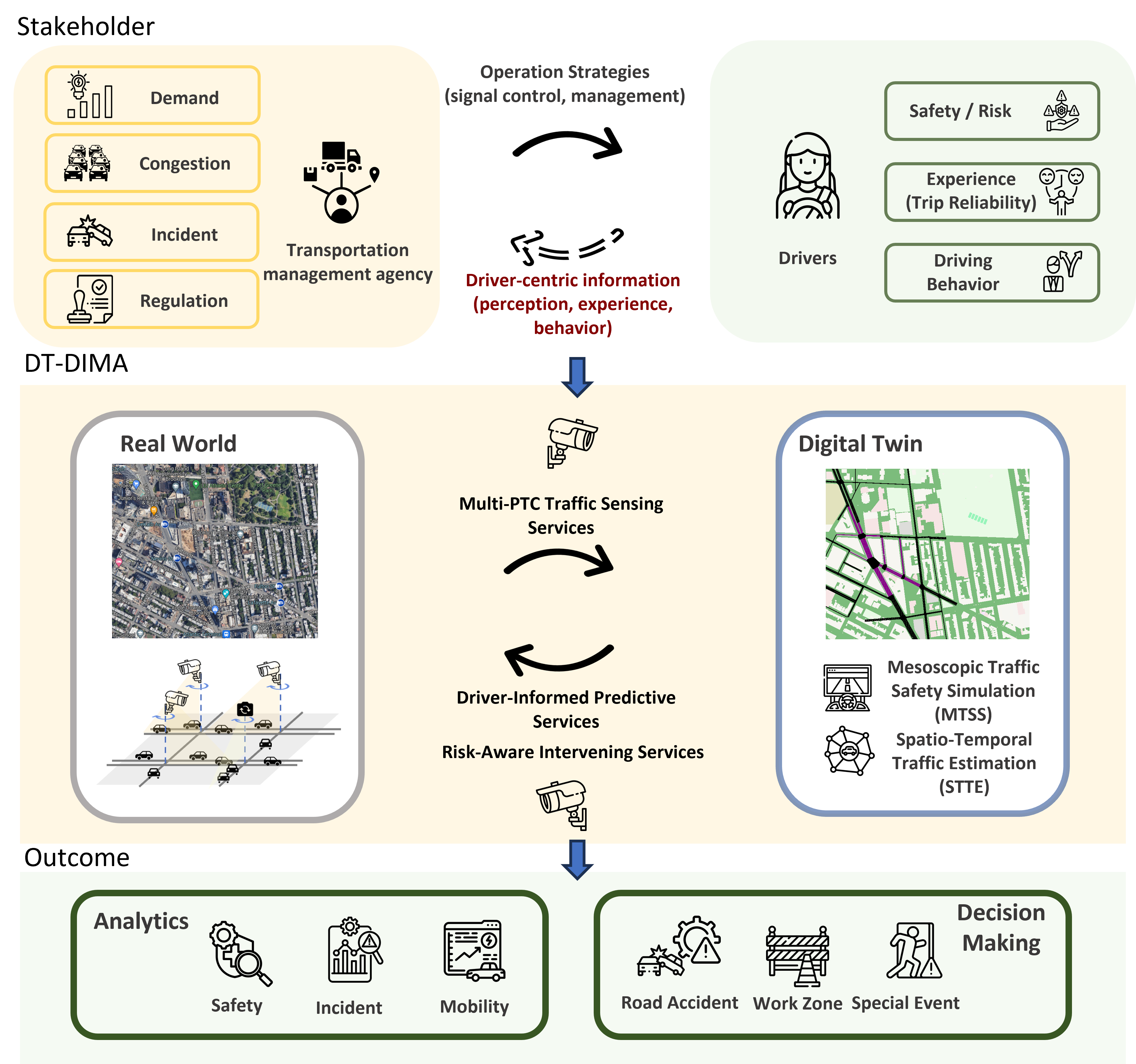}
    \caption{Feedback loops at stakeholder and modeling levels in the mobility management process}
    \label{fig:DT-DIMA concept}
\end{figure}

Urban mobility management plays a crucial role in ensuring the smooth functioning of urban transportation systems, facilitating the efficient movement of people and goods, and enhancing the overall quality of life in cities. Traditionally, urban mobility management strategies rely on bird-eye level traffic information from sensors, such as loop detectors and Wi-Fi sensors, to monitor traffic conditions and make high-level decisions \cite{traunmueller2018digital}. These strategies often prioritize risks associated with mobility, often referring to the dynamics and sudden shifts in traffic mobility patterns. However, while transportation managers emphasize the macro-level oversight gained from these sensing infrastructures, the safety risks that directly affect road users are also important. These safety risks include driving hazards and road conflicts experienced by drivers, which are critical components of overall road safety. This overlook creates a disconnection between the management perspective and the road user perspective. As shown in \Cref{fig:DT-DIMA concept}, while management operations mainly aim to enhance the overarching performance of the transportation system through measures such as congestion reduction, demand management, and incident response, the actual experiences of road users, particularly concerning safety and personal experiences, can diverge significantly from these objectives. For example, a traffic lane reconfiguration aimed at construction activities \cite{bian2019estimating} on a major urban road may improve infrastructure service from a macro perspective. However, it could increase driving risks for individual drivers as the altered traffic pattern might confuse drivers accustomed to the previous layout, leading to a rise in potential accidents \cite{garber2002distribution, yang2018methodological}. This disconnection emphasizes the importance of integrating road user perspectives into urban mobility management to ensure that strategies not only achieve their intended overarching goals in reducing mobility risks but also accommodate the needs and safety of individual road users.


The concept of a Digital Twin (DT) has emerged as an increasingly recognized technique in this context \cite{Grieves_DT} for its ability to create a virtual representation of a physical system through bidirectional communication. DT can be built as machines or computer-based models, either physical or virtual, that simulate, emulate, mirror, represent, or ``twin'' the real-world existence of a physical entity with adaptation for real-time prediction, optimization, monitoring, controlling, and assistance in improving decision-making \cite{DT_survey_2019,DT_survey_2020,DT_2022}. Leveraging input data from the physical twin, the DT is capable of predicting future states as well as enabling the simulation and testing of novel configurations, fostering proactive maintenance or management across various application domains, including but not limited to aerospace \cite{air_DT_2016,structural_DT_2017}, manufacturing \cite{Manufacturing_DT}, healthcare \cite{Hospital_DT,IoT_health_DT}, medicine \cite{Health_DT,Medicine_DT,PersonalizedMedicine_DT}, construction \cite{Civil_DT_2021,Construction_DT_2021}, and transportation \cite{Transportation_DT_survey,10380440,han2024foundation}. By integrating DT into urban mobility management, it is possible to synthesize driver-centric information and aggregate this information to network-level, thereby enhancing decision-making and aligning management strategies more closely with the needs and safety of road users. As shown in \Cref{fig:DT-DIMA concept},  the aim is to develop a DT framework that effectively communicates driver-centric information (illustrated by the dashed arrow) to transportation managers, thereby enabling more informed, driver-aware decision-making in considering not only mobility risks but also safety risks. However, the design of a DT-based framework for mobility management presents several challenges: 


\begin{itemize}
    \item \textbf{The first challenge is adapting DT for proactive management instead of being reactive to traffic incidents and disturbances.} The prevailing approach in managing traffic incidents and disturbances within urban mobility systems is predominantly reactive, focusing on addressing the aftermath rather than being proactive. This post-impact management style leads to significant response delays and congestion, as transportation managers assess the severity of incidents and implement measures to mitigate their effects \cite{10380440, lyons2018getting}. However, the inherent complexity of urban transportation networks means that this reactive strategy can exacerbate these disruptions, turning manageable incidents into hazardous scenarios such as secondary incidents that compromise road safety\cite{yang2018methodological}. Given the virtual replicas of various traffic situations in DT, the challenge lies in using DT to help mobility management address these preemptive issues.
    \item \textbf{Secondly, current vehicle-based sensing technologies fail to meet the cost-efficient implementation requirements for managing large-scale urban mobility systems using DT.} The effectiveness of a DT depends on its ability to replicate real-world traffic conditions accurately. This requires a vast amount of real-time traffic information along the road network. Connected and Autonomous Vehicles (CAVs), though equipped with advanced onboard equipment (OBE) sensing devices that can capture real-time driving situations (e.g., hard braking, acceleration, near-miss events) \cite{rahman2019safety}, have limited utility in DT due to their low penetration rate in overall transportation systems. For DT to accurately replicate driver-experienced road conflicts throughout the entire road network, a broader and more cost-efficient approach without solely relying on CAVs is necessary. 

    \item \textbf{Thirdly, a methodology for achieving real-time synergy between the physical world and digital twin in urban mobility management remains uncharted.} Effective mobility management across large-scale road networks requires operations that can respond in real time, requiring DT to synthesize real-world situations and create accurate replicas within minimal latency. These replicas must reflect the continuous evolution of the real world within the cyber-physical realm. Achieving such a high level of synergy and replication across a large-scale network requires both high fidelity and low latency. Consequently, the challenge lies in capturing real-world information swiftly, mirroring this data in digital replicas in real-time, and scaling this process across vast road networks without compromising the accuracy of the information.

\end{itemize}

To tackle the above challenges, several innovations for designing a DT-based mobility management system must be finished on both the sensors and DT. For traffic sensors implemented in the real world, enabling real-time and cost-efficient traffic sensing needs further exploration. In recent years, traffic surveillance cameras, particularly pan-tilt cameras (PTCs), have seen widespread installation and deployment along the road network. These PTCs are strategically installed at each intersection along the road network and can tilt 360 degrees to enable versatile monitoring across different areas \cite{akilan2019video, haghighat2023computer}. The researchers see the light of deploying these existing infrastructure PTCs to serve as real-time sensors by cooperatively tilting these PTCs to capture useful real-time traffic information for a large-scale network. On the DT side, the key task is to replicate the dynamics captured by the sensing services from the physical world and predict driver behaviors effectively. 

Building on the capabilities of these PTCs, a DT-based system for urban mobility management and traffic monitoring solution -- DT-based Driver risk-aware Intelligent Mobility Analytics (DT-DIMA) system is proposed. The DT-DIMA framework employs a twin-based structure comprising a physical twin representing the real world and a digital twin serving as its cyber-physical counterpart. These twins are interconnected through a feedback loop that integrates various services to enhance mobility management comprehensively. Specifically, the feedback loop incorporates three main services: the multi-PTC traffic sensing service, the driver-informed predictive service, and the risk-aware intervening service. The \textbf{multi-PTC traffic sensing service} utilizes multiple Pan-Tilt Cameras (PTCs) positioned at various intersections to collect traffic flow data, applying established video extraction technologies like multiple-object-detection (MOD)\cite{redmon2016look} and tracking \cite{wojke2017simple,Tang_2019} to derive real-time traffic metrics such as volume and speed. The \textbf{driver-informed predictive service} includes three key components. First, the Spatio-Temporal Traffic Estimation (STTE) model uses traffic data from the multi-PTC sensing service to provide estimated network-wide traffic states. Second, the Mesoscopic Traffic Safety Simulation (MTSS) model \cite{sha2023calibrating} simulates individual driver behaviors and assesses driving risks under various traffic conditions, overcoming the challenge of low Connected and Autonomous Vehicles (CAVs) penetration, which limits transportation managers for gaining access to driver-centric data. Lastly, the Long-Short Term Twinning (LSTT) algorithm monitors the performance of MTSS and adjusts the update frequency of the model based on its accuracy and reliability. The \textbf{risk-aware intervening service} implements the Risk-Constrained Correlated Online Learning (RiCCOL) approach, which leverages safety risks and mobility risks from the predictive service to optimize the tilting strategies of PTCs. RiCCOL will control and tilt PTCs to those road segments with high driving risks and abrupt changes in mobility patterns.

With the proposed system, two folds of risk information can be unveiled: mobility risk and safety risk. The mobility risk refers to the dynamics and abrupt changes residing in the traffic mobility patterns, including traffic flow and speed at both edge-level and network-level. The safety risk refers to the driving risks experienced by road users, the predicted driving risks are also aggregated to edge-level as well as network-scale. The performance of the DT-DIMA system is evaluated using a simulation dataset calibrated using real-world data from Brooklyn, New York. Extensive experiments are conducted under various traffic situations, such as unexpected accidents and pre-planned construction events. Additionally, a case study showed that the DT-DIMA system can enhance decision-making by projecting safety outcomes of various strategies without the costs of actual implementation. This allows transportation managers to foresee safety risks and proactively allocate resources to mitigate negative impacts on drivers. In summary, the contributions are summarized as follows:



\begin{itemize}
    \item We propose DT-DIMA, a novel DT-based system using pan-tilt cameras (PTCs) to replicate real-world traffic dynamics within urban mobility systems. This system seamlessly mirrors diverse traffic scenarios and evaluates the safety risks faced by drivers, providing transportation managers with real-time insights into the efficacy of their traffic management strategies. By integrating predictive analytics, DT-DIMA offers actionable intelligence to enhance traffic safety and optimize management decisions. 
    \item To effectively gather driving risk information from the transportation networks, a novel driver-informed predictive service is developed based on the DT structure. Extensive experiments on a large-scale network show that our proposed service can predict driving risks encountered by drivers and achieve MAPE ranging from $0.85\%$ to $12.97\%$ across normal conditions, pre-planned work zones, and unexpected accident scenarios.
    \item To enable the collaborative operation of PTCs along the road network, a Risk-Constrained Correlated Online Learning (RiCCOL) algorithm is designed in a risk-aware intervening framework, allowing PTCs to respond seamlessly to hazardous road segments with high driving risks.  Thanks to the safety risk prediction by MTSS, RiCCOL can preemptively tilt PTCs to cover edges with high driving risks before incidents occur. Our experiments demonstrate that PTCs equipped with the RiCCOL algorithm can effectively monitor around $70\%$ of hazardous roads and create around a 5-minute lead time in capturing incidents. 
    \item To optimize the use of computational resources while maintaining an accurate reflection of the physical world, an innovative algorithm named adaptive Long Short-Term Twinning (LSTT) is designed. This approach empowers the DT-DIMA system to adaptively alternate between long-term and short-term simulations within the MTSS. Ablation experiments on LSTT in reaction to situations like changes in traffic demands as well as unexpected accidents showed that it can adaptively detect such changes and improve driving risk prediction by $2\%$ MAPE when the accident is in effect.  
\end{itemize}

\section{Literature Review}
\subsection{Review of DT in general sectors}
In aviation applications, DT often functions as a predictive maintenance tool to rapidly identify and address potentially critical changes in the structural integrity of aircraft, such as detecting fatigue cracks \cite{air_DT_2016}. DT then triggers self-healing mechanisms as a responsive action. Moreover, DT can contribute to decision support, optimization processes, and diagnostic analyses for the aircraft \cite{structural_DT_2017}. 
Within the manufacturing context, DT is utilized to optimize all aspects of the product manufacturing process, including supervising each step of manufacturing, identifying potential failures of machines, and finding the optimal management solution with self-adaptation to its physical twin \cite{Manufacturing_DT,DT_survey_2019}.
In the healthcare sector, DT can be utilized in many different ways \cite{IoT_health_DT}, from predictive maintenance of medical devices and device performance optimization to hospital management \cite{Hospital_DT}. 
Regarding medical and clinical usages, the interest in DT is driven by the aspiration to create an organ (e.g., heart, airway system, etc.) or a human DT \cite{Medicine_DT}. The DT aims to represent the internal state of its physical counterpart, facilitating the prediction of potential illnesses by analyzing the real twin's personal health history and other related features \cite{Health_DT}. Then, more personalized medicine \cite{PersonalizedMedicine_DT} that targets the needs of the patients based on their own genetic, biomarker, phenotypic, physical, or psychosocial characteristics can be utilized for planning treatment that better fits the patient.
In the construction industry, DT can address the shortcomings of Building Information Modeling (BIM) in facilities management by providing real-time updates on a building's status post-commissioning \cite{BIM_DT}. DT also offers a holistic and integrated approach, ensuring effective construction monitoring and control by combining information from multiple systems \cite{Build_Monitor_DT}.

\subsection{Review of DT in transportation sectors}
For applications in the smart transportation sector, DT is essential in elevating both efficiency and safety. Recent research efforts can be broadly categorized into two major trends: a rather detailed-level DT that involves the analysis of vehicle trajectories, exploration of road users' behavior, and simulation of traffic environments for enhancing the safety of autonomous vehicles (AVs); a broader-scale DT that aims at assisting the Transportation Management Center (TMC) in tasks such as traffic management, transportation planning, and safety analyses over the network. 
In the former, well-noted works such as the traffic environment simulator TeraSim leverage DT at the infrastructure level and employ dense deep-reinforcement learning for unbiased and efficient AV safety testing \cite{feng2021intelligent, feng2023dense}. Experimental DT, as studied by \cite{keler2023calibration}, assists in proving the importance of calibrating microscopic traffic simulations and testing the reliability of driving simulators. Additionally, \cite{Zheng_2023} collects the CitySim dataset of vehicle trajectories by Unmanned Aerial Vehicles (UAVs), which can be used in modeling DT for safety research. 
In the latter, for traffic monitoring, \cite{road_DT} deploys Digital Twin Boxes to roads, continuously sending data to the edge or cloud for constructing DT representations of physical roads. In the context of traffic data acquisition, \cite{azfar2023incorporating} establishes a campus DT and visualizes detected and tracked physical vehicles in the cyber twin. For prediction, \cite{trans_pred_DT} analyzes data within the DT, incorporating flow and velocity data from Internet of Vehicle (IoV) sensors to predict traffic flow. For traffic management purposes, \cite{Sec_ITS_DT} adopts DT and deep learning algorithms to address problems such as traffic assignment and signal control by establishing online and offline modules in the virtual twin, and investigates potential security issues in intelligent transportation systems. Our study aligns with the broader-scale DT, incorporating capabilities in data acquisition, traffic state forecasting, and safety-driven event monitoring that can assist the TMC in making management decisions, including but not limited to camera and traffic signal control, congestion management, and evacuation.

It is important to emphasize that the complete benefits of DT in intelligent transportation systems are yet to be fully acknowledged and realized. According to Transportation 5.0 mentioned in \cite{10380440,han2024foundation}, DT must go beyond vehicular movements and extend the scope to encompass broader transportation systems, network dynamics, and human decision-making processes.

\section{Methodology}
\subsection{Architecture of DT-DIMA system}

\begin{figure}[t!]
    \centering
    \includegraphics[width=1.\columnwidth]{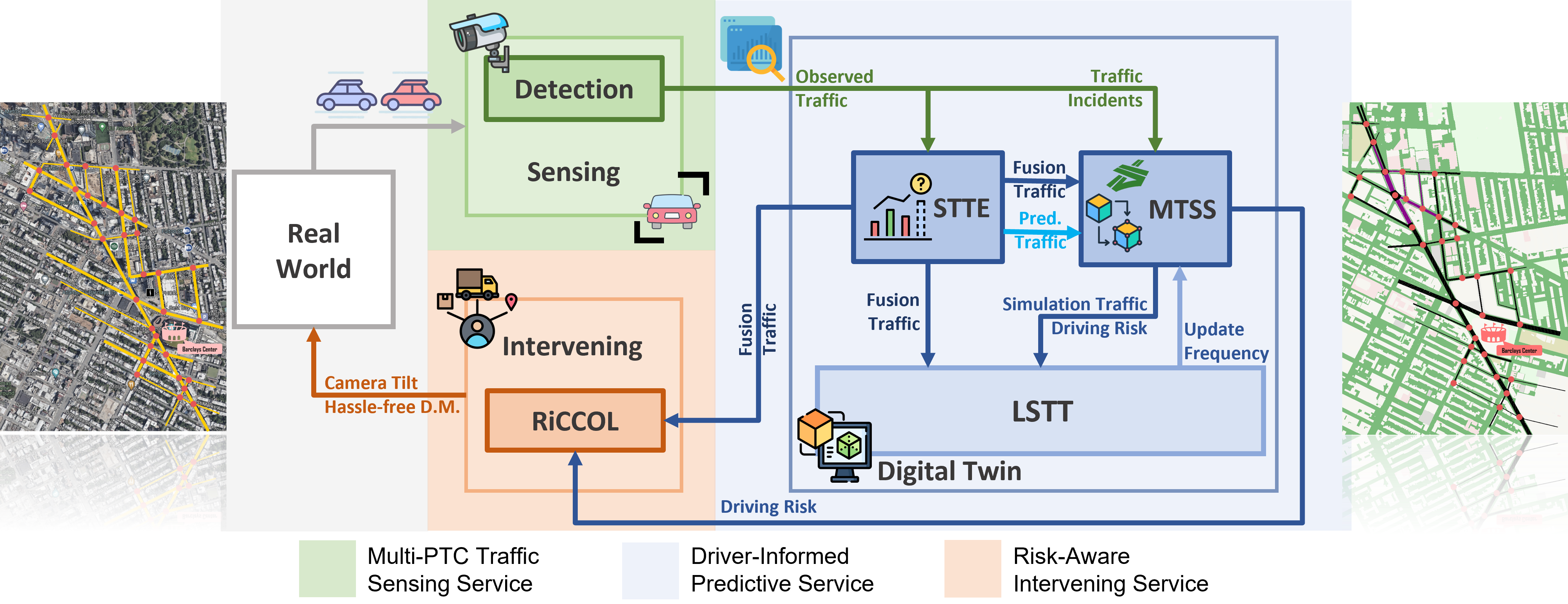}
    \caption{Architecture of overall framework of DT-DIMA. Major components in this figure, STTE: Spatio-Temporal Traffic Estimation, MTSS: Mesoscopic Traffic Safety Simulation, LSTT: Long-Short Term Twinning, RiCCOL: Risk-Constrained Correlated Online Learning.}
    \label{fig:DT-DIMA architect}
\end{figure}
The DT-DIMA system integrates three key services to synchronize with the real-world transportation mobility system: the multi-PTC traffic sensing service, the driver-informed predictive service, and the risk-aware intervening service. The architecture of the DT-DIMA system, as depicted in \Cref{fig:DT-DIMA architect}, begins with the multi-PTC traffic sensing service capturing data from physical transportation systems. This includes traffic states along road segments and traffic incidents such as accidents and work zones. This gathered data is processed by the driver-informed predictive service, which creates a DT with three components: STTE, MTSS, and LSTT. The DT creates a real-time replica of the physical world using the gathered data and provides real-time predicted traffic states as well as driving risks to the risk-aware intervening service.

Building on these predictions, the risk-aware intervening service develops real-time decision-making strategies for both surveillance and mobility management. The RiCCOL algorithm uses the fused traffic data from STTE and driving risks from MTSS to guide the real-time tilting of PTCs across the road network. The intermediate variables and parameters are summarized in\Cref{tab:notations}.
\begin{table}[!h]
    \centering
    \begin{tabular}{ll}
    \toprule
    Notation(s)  & Description\\
    \midrule
    $\mathcal{G}$, $\mathcal{N}$, $\mathcal{E}$  & The traffic network, the junction (nodes) set, and road segments (edges) set  \\
    $(i,j)\in \mathcal{E}$, $i,j\in \mathcal{N}$ & The directed edge from node $i$ to $j$\\
       $T$, $[T]:=\{1,2,\dots, T\}$  & The horizon length and the set of discrete time indices  \\
       $\mathcal{N}^c$ & The subset of nodes where cameras are deployed\\
       $\mathcal{A}^{i}$, $a^i_t$ & The action set of camera $i$ and the implemented action at time $t$\\
       $\bm{a}_t$ & The joint tilting action at time $t$\\ 
       $\bm{s}_t$, $\bm{s}_t(i,j)$ &  The network-level traffic state and its $(i,j)$-entry (the edge-level traffic state)  at time $t$\\
       $\bm{q}_t$, $\bm{q}_t(i,j)$  & The network-level and edge-level traffic volume at time $t$\\
       $\bm{v}_t$, $\bm{v}_t(i,j)$  & The network-level and edge-level traffic speed at time $t$\\
       $\bm{\delta}_t$, $\bm{\delta}_t(i,j)$ & The network-level and edge-level relative traffic state fluctuation at time $t$, respectively\\
       $\bm{s}_t^c $ & The joint observation of the PTCs\\
       $\bm{\delta}_t^f$ &  The traffic  fluctuation distribution calculated using fusion data\\
       $\hat{\bm{s}}_t$ & The  traffic state prediction returned by the STTE\\
       ${\bm{s}}_t^f, \bm{q}_
t^f, \bm{v}_t^f$ & The traffic state fusion (volume and speed fusion), a combination of $\hat{\bm{s}}_t$ and $\bm{s}_t^c $\\
       $\bm{r}_t$ & The actual SSM distribution over the network\\
       $\hat{\bm{r}}_t$ & The predicted SSM distribution over the network produced by the SUMO\\ 
       $\hat{\bm{q}}^*_t, \hat{\bm{q}}^*_t(i,j)$ & The network-level and edge-level traffic volume output by MTSS\\
       $\hat{\bm{v}}^*_t, \hat{\bm{v}}^*_t(i,j)$ & The network-level and edge-level traffic speed output by MTSS\\
       $\bm{Z}_{t} = ({Z}^{ij}_t)$ & The traffic incident information, including the edge $(i,j)$\\ &where the incident locates, the time $t$ during the occurrence of incident\\
       $P^*$ & The simulation period, a period in minutes between two calls of simulation\\
    \bottomrule   
    \end{tabular}
    \caption{A summary of frequently used notations.}   
    \label{tab:notations}
\end{table}

\subsection{Multi-PTC traffic sensing service}
In the proposed DT-DIMA system, the multi-PTC traffic sensing service is mainly powered by computer vision-based deep learning methods. Specifically, it utilizes object detection methods such as YOLO series\cite{bochkovskiy2020yolov4} and object tracking methods like StrongSORT\cite{du2023strongsort}. These models process video frames captured by multiple PTCs, which are strategically positioned at each intersection throughout the transportation network.
By integrating object detection and tracking methods, the PTCs function as real-time traffic sensors, generating essential traffic data such as vehicle counts, speeds, and traffic incidents along their tilt directions. As shown in \Cref{fig:DT-DIMA architect}, this data forms the foundation for the driver-informed predictive service, enabling the Digital Twin (DT) to mirror real-world traffic conditions accurately and in real time. However, since a PTC can only tilt in one direction at a time, relying solely on data from individual PTCs is insufficient for constructing comprehensive network-scale traffic information. Thus, the multi-PTC sensing service must collaborate with the other two services within the DT-DIMA system to ensure a complete capture and analysis of traffic data across the large-scale road network.

In this study, we consider a traffic network represented by a graph $\mathcal{G}=(\mathcal{N}, \mathcal{E})$, where $\mathcal{N}$ denotes the set of road intersections, and $\mathcal{E}$ denotes the set of the road segments (edges). PTCs are assigned to a subset of intersections for surveillance purposes. Let $\mathcal{N}^c\subset \mathcal{N}$ be the set of the nodes equipped with a PTC. With a slight abuse of notation, we denote by $i\in \mathcal{N}^c$ the PTC deployed at the $i$-th intersection, which can be tilted to monitor the inbound and outbound links $(i,j)$ and $(j, i)$ from a neighboring node $j$ at a time. Each camera's action is defined as a $\mathcal{E}$-dimensional binary vector $a_t^i\in \mathcal{A}^i:=\{0, 1\}^{|\mathcal{E}|}$.  The one entries correspond to the monitored edges: $a_t^i{(i,j)}=1$, if $(i,j)$ is covered by the PTC at time $t$. Let $\bm{a}_t=\vee_{i\in \mathcal{N}^c} a_t^i\in \{0,1\}^{|\mathcal{E}|}$ be the joint tilting action of all cameras at time $t$, where $\vee$ denotes the entry-wise boolean operator ``or''. In plain words, $\bm{a}_t(i,j)=1$ if the edge $(i,j)$ is covered by at least one of the PTCs.

We model the traffic evolution within the network during a given period of time using a discrete-time sequence $\{\bm{s}_t\}_{t\in [T]}$, where $\bm{s}_t$ denotes the real-time traffic information of interest at $t$-th minute. The traffic state variable $\bm{s}_t\in \R^{|\mathcal{E}|}$ denotes the aggregated local information from each edge $\bm{s}_t(i,j)$. Depending on the use cases, the traffic state $\bm{s}_t(i,j)$ can be the edge-level traffic volume (the number of vehicles passing through a particular segment), denoted by $\bm{q}_t(i,j)$, or the edge-level average speed, denoted by $\bm{v}_t(i,j)$. 

For simplicity, we use $\bm{s}_t$ as a generic notation for the traffic state and introduce the following helpful notations, which can be similarly defined for the traffic volume $\bm{q}_t$ and speed $\bm{v}_t$.  Employing the entry-wise product (Hadamard product), denoted by $\otimes$, the joint observation of the multi-PTC service is given by $\bm{s}_t^c=\bm{s}_t\otimes \bm{a}_t$, where the edge-level traffic states without any camera monitoring are masked with zero. Additionally, PTCs can also extract traffic incident information at each time step across the network, denoted by $\bm{Z}_t$, where $\bm{Z}_t(i,j)$ contains incident information on the edge $(i,j)$, e.g., incident location, affected lanes, and is an empty set if no incident happens at time $t$.

\begin{figure}[t!]
    \centering
    \includegraphics[width=1.\columnwidth]{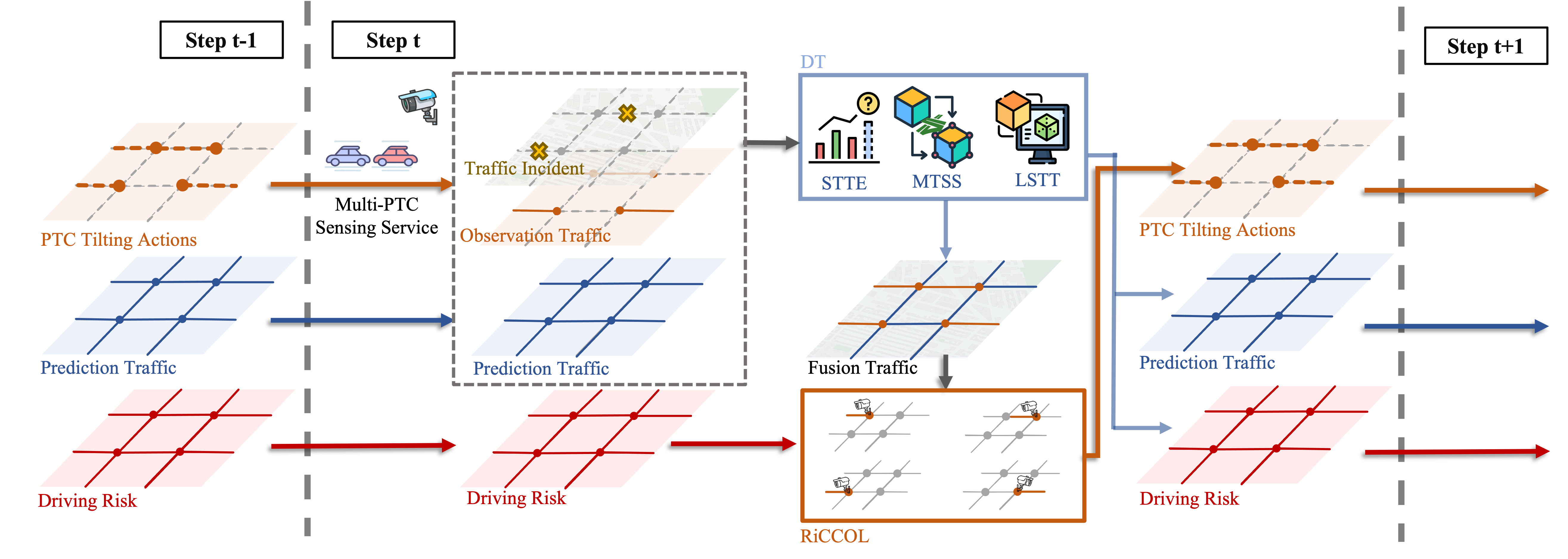}
    \caption{Workflow of DT-DIMA in real-time operation}
    \label{fig:DT-DIMA workflow}
\end{figure}

\subsection{Driver-informed predictive service}
\begin{figure}[t!]
    \centering
    \includegraphics[width=1.\columnwidth]{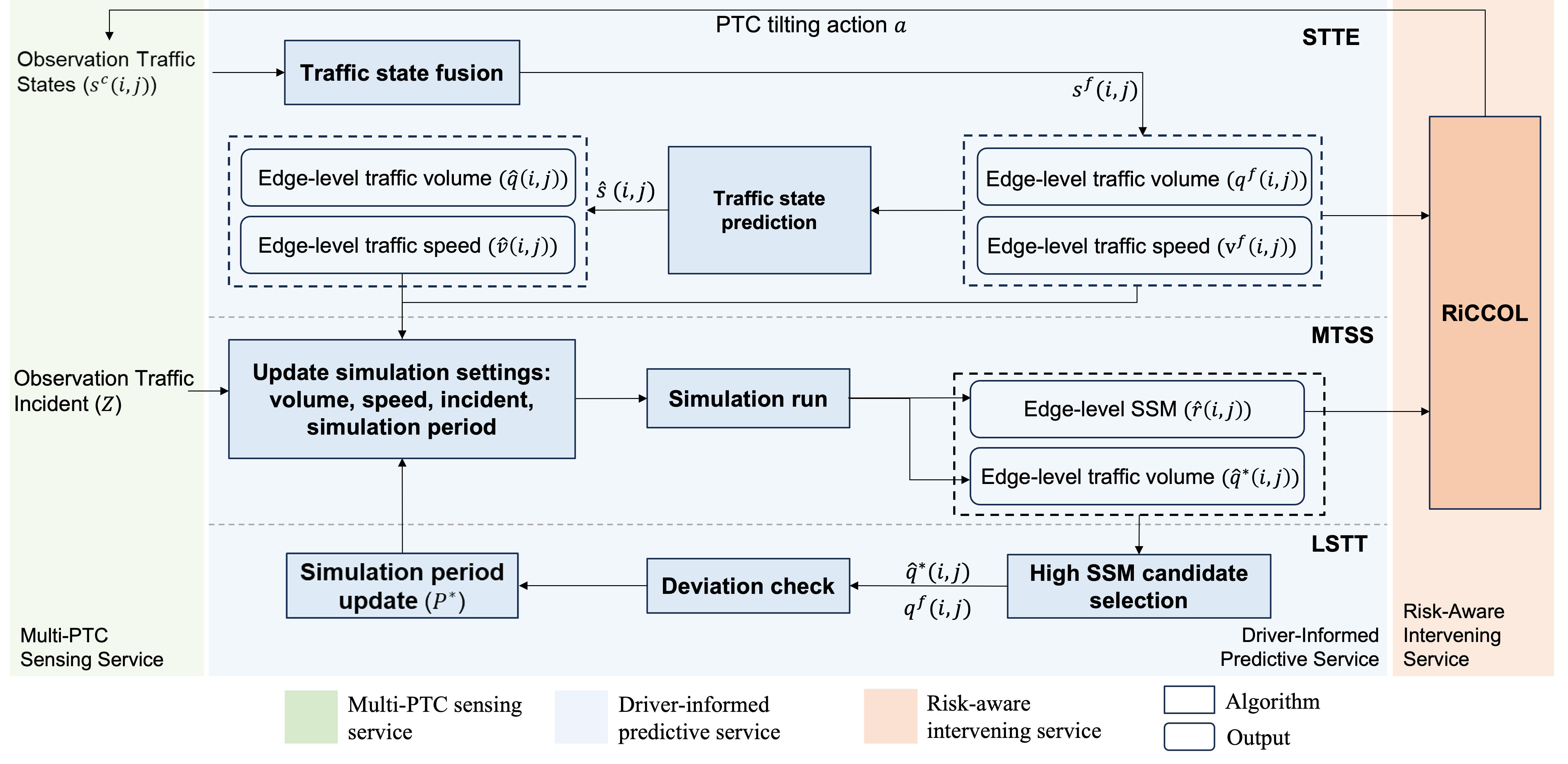}
    \caption{Workflow of driver-informed predictive service}
    \label{fig:DT-workflow}
\end{figure}

The workflow of the proposed driver-informed predictive service is detailed in \Cref{fig:DT-workflow}. Initially, the STTE component receives observed traffic states from the multi-PTC sensing service. It processes these observations to generate fused traffic states and predicted traffic states. Both the fused and predicted traffic states are then utilized as inputs for the MTSS component, while the fused traffic states are also used by the RiCCOL algorithm within the risk-aware intervening service.

The MTSS component integrates this data with observed traffic incident information, also sourced from the multi-PTC sensing service. It uses the fused traffic states, including traffic volume and speed from the STTE, to update its simulations for the current stage and employs the predicted traffic states to guide future traffic demand. These simulations generate predicted outputs such as driving risks. Subsequently, the LSTT component uses the simulated traffic volume from MTSS and the observed traffic volume from STTE to determine the necessity of updating the simulation period.

\subsubsection{Spatio-temporal traffic estimation model (STTE)}
The motivation for introducing STTE into driver-informed predictive service in DT structure is to capture the network-level traffic information and aid MTSS in more accurately replicating the physical world. In the multi-PTC sensing service, the traffic information can only be captured by a PTC at one direction at a time, which inherently restricts the comprehensive coverage across an extensive road network. This limitation highlights the need for the STTE to provide accurate traffic estimations for the directions and areas that are not directly observed.

\begin{figure}[t!]
    \centering
    \includegraphics[width=1.\columnwidth]{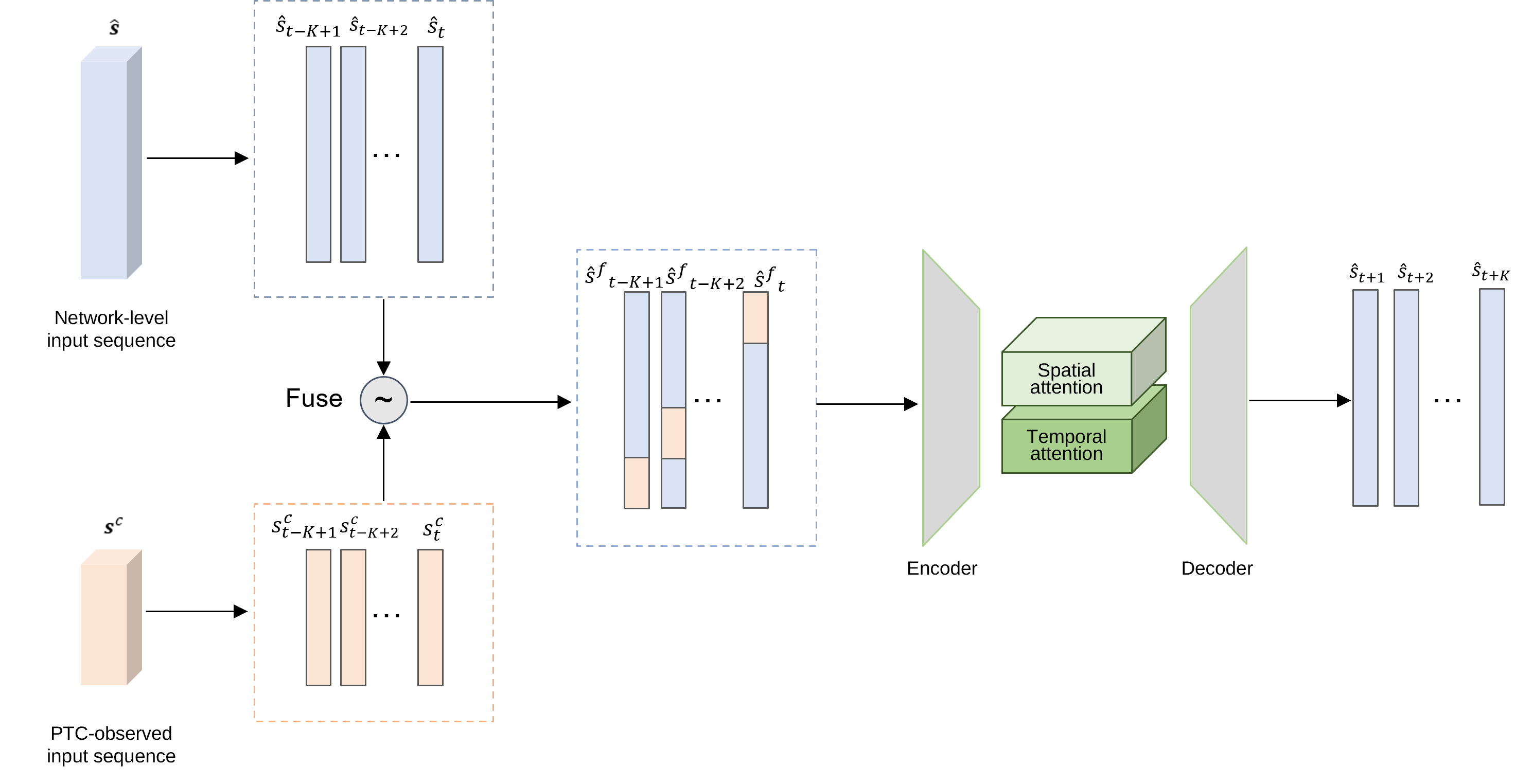}
    \caption{The spatio-temporal traffic estimation (STTE) component}
    \label{fig:STTE component}
\end{figure}

\paragraph{Problem formulation of STTE} 
As shown in \Cref{fig:STTE component}, given a sequence of recent traffic state observations ${\{\bm{s}_\tau^c\}_{\tau=t-K+1}^{t}}$ over the past $K$ time steps, STTE aims to predict the sequence of future traffic states over the next $K$ time steps. To this end, STTE first fuses the recent observations with the predictions: $\bm{s}_\tau^f=\bm{s}_\tau^c+\hat{\bm{s}}_\tau\otimes(\mathds{1}-\bm{a}_\tau)$, $\tau\in [t-K+1, t]$, where $\mathds{1}\in \R^{|\mathcal{E}|}$ is an all-one vector. Then, STTE takes in this fusion data as input and predicts the traffic states in future $K$ steps, denoted by $\{\hat{\bm{s}}_\tau\}_{\tau=t+1}^{t+K}$. This fused traffic state $\{\bm{s}_t^f\}$ also serves as input to the MTSS component and LSTT component. 

\paragraph{Local attention kernel}
The aim of STTE is to predict traffic states of edges that are not directly observed by PTCs, which presses the need to capture local information. Specifically, at timestamp $t$, node $i$ ($i$th intersection) receives observation $s_t(i,j)$ for edge $(i,j)$, there is a need to estimate $\hat{s}_t$ along the rest of edges that directly connect to node $i$. Furthermore, the traffic states along these edges are also influenced by their immediate connected edges in upstream and downstream directions.
To capture this locality, we design a local attention kernel using the graph convolution in Graph Convolution Network (GCN) \cite{kipf2016semi}. The graph convolution propagates messages along a local view, ensuring the message passes to the immediate neighbors. The graph convolution can be formulated as follows:
\begin{equation}
\label{Eq: GCN}
    \begin{aligned}
       \text{GCN}(H^{(l)}, \Tilde{A}) = \sigma \Big( \big((D^{-1} \Tilde{A})\odot \mathbf{S} \big) H^{(l)} W^{(l)} \Big).
    \end{aligned}
\end{equation}
where $H^{(l)} \in \mathbb{R}^{|\mathcal{E}| \times d_{\text{model}}}$ is the graph signal representation in the $l^{th}$ layer. $\Tilde{A} = A + I$ is the adjacency matrix of the directed graph $G$ with added self-connections: $I \in \mathbb{R}^{|\mathcal{E}|\times|\mathcal{E}|}$ is the identity matrix. It is important to note that in this study, matrix $A\in \R^{|\mathcal{E}|\times|\mathcal{E}|}$ describes the connectivity between edges $(i,j)\in \mathcal{E}$, rather than between nodes. $D$ is degree matrix and $W^{(l)}$ is layer-specific trainable weights. $\sigma(\cdot)$ is the nonlinear activation function. 

Unlike traditional GCN models that use a static adjacency matrix $\Tilde{A}$ to represent local spatial correlations, this paper incorporates a dynamically updating spatial correlation weight matrix $\mathbf{S}$ \cite{guo2021learning}:
\begin{equation}
\label{Eq: SA matrix}
    \begin{aligned}
       \mathbf{S} = \text{softmax} \Big(\frac{H^{(l)}H^{(l)\top}}{\sqrt{d_{\text{model}}}}\Big) \in \mathbb{R}^{|\mathcal{E}| \times |\mathcal{E}|}.
    \end{aligned}
\end{equation}
Spatial correlation matrix $S$ learns local attention intensities dynamically, allowing the local attention kernel to adjust in response to changing traffic conditions along the local neighborhood around each intersection.

\paragraph{Global attention kernel}
Graph convolution effectively captures local or short-range spatial dependencies within traffic patterns, but the complexity of traffic dynamics also involves long-range interactions that can influence broader areas of the road network. Unforeseen events, such as accidents or road closures, can trigger cascading effects that ripple through the entire network, leading to widespread congestion and disruptions far from the original incident site. This paper employs the attention mechanism of the encoder-decoder structured transformer, as described in\cite{vaswani2017attention}, which facilitates direct communication between distant road segments. By attending to all positions simultaneously, the multi-head attention component of the transformer can effectively model long-range dependencies in traffic dynamics. The multi-head attention can be expressed as:
\begin{equation}
\label{Eq: multi-head}
    \begin{aligned}
       \text{MultiHead}(\mathbf{Q}, \mathbf{K}, \mathbf{V}) & = \oplus \big(\text{head}_1, \cdots, \text{head}_h\big) W^O,\\
      \text{head}_i & = \text{Attention}\big(\mathbf{Q}W^Q_i, \mathbf{K}W^K_i, \mathbf{V}W^V_i\big),
    \end{aligned}
\end{equation}
where the query $Q \in \mathbb{R} ^{L_Q \times d_{\text{model}}}$, key $K\in \mathbb{R} ^{L_K \times d_{\text{model}}}$ and value $V \in \mathbb{R} ^{L_V\times d_{\text{model}}}$ are both transformed from the input token sequence $X \in \mathbb{R} ^{L \times C}$, and $L_Q, L_K, L_V$ are the corresponding length. Linear projection $W_i^Q, W_i^K, W_i^V \in \mathbb{R}^{d_{\text{model}} \times d_k}$ map dimension of $d_{\text{model}} $ into head $h_i$.

As shown in \Cref{fig:STTE component}, by combining the local attention kernel and global attention kernel, we assemble the long and short-range traffic propagation parallel in each layer. The interplay of global and local attention kernels ensures STTE accurately estimates unobserved traffic states for individual PTC and captures the network-level traffic dynamics within mobility patterns.

\subsubsection{Mesoscopic traffic safety simulation model (MTSS)}\label{subsubsec: MTSS}
The MTSS component is designed to accurately replicate various driving behaviors and simulate the associated risks for each vehicle. This component is developed using a mesoscopic simulator called the Simulation of Urban Mobility (SUMO) \cite{SUMO2018}. SUMO is an advanced, open-source traffic simulation tool that handles extensive networks and offers detailed vehicular movement simulations. It models individual vehicles, allowing for the observation of complex interactions and behaviors typical in real-world traffic. This fidelity is crucial for accurately predicting traffic dynamics, particularly when responding to disruptions or variations in standard traffic flows. 

Leveraging SUMO, the MTSS component integrates real-time traffic information from the STTE component as well as the multi-PTC sensing service. It uses this comprehensive data to create a realistic and timely updated simulation of the traffic environment, mirroring the physical world within the SUMO framework. This simulation enables MTSS to accurately replicate traffic conditions and the driving risks encountered by individual drivers.

\paragraph{Calculation of Time-to-Collision (TTC) events in MTSS} 
In MTSS, Time-to-Collision (TTC) is defined for all follow-lead scenarios where the following vehicle is faster than the leading vehicle, calculated as:
\begin{equation}
\label{Eq: TTC_1}
    \begin{aligned}
       \mathbf{TTC} = \frac{\text{Space Gap}}{\text{Speed Difference}}
    \end{aligned}
\end{equation}
For crossing or merging situations, TTC is considered if the expected conflict area exit time for vehicle A (vehicle entering the conflict area first) is greater than the entry time for vehicle B. Under these conditions, TTC is calculated as:
\begin{equation}
\label{Eq: TTC_2}
    \begin{aligned}
       \mathbf{TTC} = \frac{\text{Distance to Conflict Area Entry of Vehicle B}}{\text{Current Speed of Vehicle B}}
    \end{aligned}
\end{equation}

In the MTSS traffic simulation, three primary vehicle conflict types have been identified: Lead/Follow, Crossing, and Merging situations. In the Lead/Follow situation, vehicles maintain the same lane sequence both prior to and following the conflict point. Conversely, in Crossing situations, vehicles navigate different lane sequences before and after the conflict point, whereas in Merging situations, vehicles originate from different lanes but converge into the same lane post-conflict. Each conflict type is assigned a designated entry point, and for Crossing and Merging situations, an exit point is also defined. These points differ slightly among vehicles as they are determined by the positions of the front or rear bumpers, although collisions may involve various parts of the vehicles. Due to software constraints, the geometry of the conflict areas is simplified; crossing lanes are assumed to be orthogonal, and merging areas are not distinctly modeled but are treated as the intersection at the start of the shared target lane. Importantly, these conflict scenarios are not merely theoretical constructs but are derived from real drivers' experiences. The framework leverages data collected from actual driving incidents, ensuring that the modeled situations reflect the complex and varied nature of real-world driving behaviors and conflicts. This research utilizes a 3-second threshold for the Time to Collision (TTC), focusing on the minimum TTC (minTTC, both time and position) as the key event data according to the previous traffic safety research \cite{sha2023calibrating,wang2021review}. 

\paragraph{Real-time update of simulations} 
To ensure that the simulations within the MTSS closely replicate actual traffic conditions, input parameters are meticulously derived from real-world observations as well as STTE. During each simulation call in the MTSS, three key pieces of traffic information are updated: traffic flow derived from the average volume on each edge, traffic speed variations on each edge, and traffic incident information. This updating is executed by Calibrators set on every edge, which dynamically adapt traffic flows, speeds, and vehicle parameters based on real-time measurements. Calibrators in SUMO are trigger-type objects specified within an additional file, enabling real-time adjustments of traffic flows, speeds, and vehicle parameters. Specifically, the update process occurs in two stages:
\begin{itemize}
    \item \textbf{Updates of current stage:} The MTSS first updates to reflect the most current traffic information since the last simulation call. Traffic flow is updated based on the average traffic volumes on the each edge, using the fused traffic volume data ($\bm{q}^{f}(i,j)$) from the STTE component. Similarly, the speed distribution is updated using the fused speed data ($\bm{v}^{f}(i,j)$) from the STTE. In the event of traffic incidents, the simulation configures the impact on specific lanes based on the observed incident's location and time period from the previous $P^*$ minutes. This incident data ($\bm{Z}$) sets a temporal window for adjusting the simulation parameters accordingly.
    \item \textbf{Updates of future stage:} Once the MTSS has synchronized with the latest physical traffic conditions, it uses the predicted traffic volume and speed on each edge from the STTE to guide the future distribution of traffic demands within the network in the future $P^*$ minutes. 
    

\end{itemize}

The MTSS then predicts driving risks along each road segment based on these updated traffic scenarios, ensuring that the simulations provide accurate foresight into potential safety risks.

\paragraph{Outputs of MTSS}
The outputs generated from our MTSS simulations are detailed and structured into two primary categories to support LSTT and RiCCOL modules work.

\noindent \textbf{Traffic State Information:}
The simulation provides extensive general traffic data for each edge, updated minute-by-minute. This output includes edge-level average traffic volume, speed, and density. To differentiate from those outputs by STTE, we denote by $\hat{\bm{q}}^{*}_{t}(i,j)$ and $\hat{\bm{v}}^{*}_{t}(i,j)$ the edge-level traffic volume and speed output by MTSS, respectively. These simulation data are directly used in the LSTT component to evaluate the effectiveness of the MTSS module, which will be introduced in the following subsection.

\noindent \textbf{Surrogate Safety Measure (SSM) Events:}
The SSM outputs encapsulate critical details of potential conflict events within the traffic network. For each identified event, the simulation records the time at which the event (or near-miss) was closest to occurring (minTTC recording time), providing a precise temporal context. Additionally, it captures the specific location within the network where the conflict occurred, which is essential for spatial analysis of high-risk areas.
We use the density of SSM events (in short, SSM) per $ d_t$-minutes on each edge to quantitatively depict the real-time driving risk. Denote by $ssm_t(i,j)$ the number of SSM events on edge $(i,j)$. For the $k$-the $ d_t$-minute interval, the SSM event density is defined as 
\begin{equation}
    \label{Eq: SSM_Count}
    \hat{\bm{r}}_{k}(i,j)= \frac{\sum_{t=k d_t}^{(k+1) d_t-1} ssm_t(i,j)}{\sum_{t=k d_t}^{(k+1) d_t-1}\hat{q}^*_t(i,j)},
\end{equation}
where we use the convention that $0/0=0$ if the accumulative volume is zero, i.e., no vehicle running on the edge. An appropriate interval length $d_t\in \N_{+}$ should cover a complete signal cycle and travel time of an average vehicle so that the SSM metric can properly reflect the driving risks users experience on the corresponding edge. Our experiments set the interval to 5 minutes.  The SSM serves as the key input to LSTT and RiCCOL modules (see \Cref{fig:DT-DIMA workflow}), and hence, we expand the SSM sequence $\{\hat{\bm{r}}_{k}\}_{k\in [T/d_t]}$ to the whole horizon as RiCCOL runs on a minute-by-minute basis. Let $\hat{\bm{r}}_{t}\triangleq \hat{\bm{r}}_{k}$, if $t\in [k d_t, (k+1)d_t)$, and we refer to $\hat{\bm{r}}_{t}$ as the network-wide SSM distribution. For simplicity purposes, we use the term SSM to represent driving/safety risks in the remainder of this paper.


\subsubsection{Long short-term twinning (LSTT)}
Since MTSS is built upon an agent-based simulation platform, one instance of simulation consumes a substantial amount of computation resources. Therefore, unlike STTE, calling MTSS on a minute-by-minute basis is computationally costly when deployed online. Instead, we use periodic execution: MTSS simulation is called every $P^*$ minutes to simulate traffic dynamics in future $P^*$ minutes. Such a simulation period is instrumental in the accuracy and realism of MTSS simulation, which leads to a trade-off between simulation cost efficiency and fidelity. To simplify our exposition below, we consider two kinds of simulation periods $P^*$, 60 minutes and 30 minutes, as representatives of the long and short-term simulation, respectively.     

Within the same service horizon, e.g., 24 hours, it is straightforward to see that the long-term simulation leads to fewer updates of MTSS and less computational expense. However, the downside is that long-term simulation may produce outdated replicas that lag behind real-world conditions, potentially misleading downstream services. {To see this, recall that the MTSS sets the configuration based on past incident information collected by PTCs in the previous $P^*$ minutes, to simulate traffic states in the future $P^*$ minutes.} The larger the period $P^*$ is, the fewer updates of MTSS, the more MTSS's replica lags behind the real world. For example, should a traffic incident happen right after one call of MTSS on a 60-minute period, it will not appear in the DT until the next call, by which one hour has elapsed.  
 
\begin{figure}[t!]
    \centering
    \includegraphics[width=0.6\columnwidth]{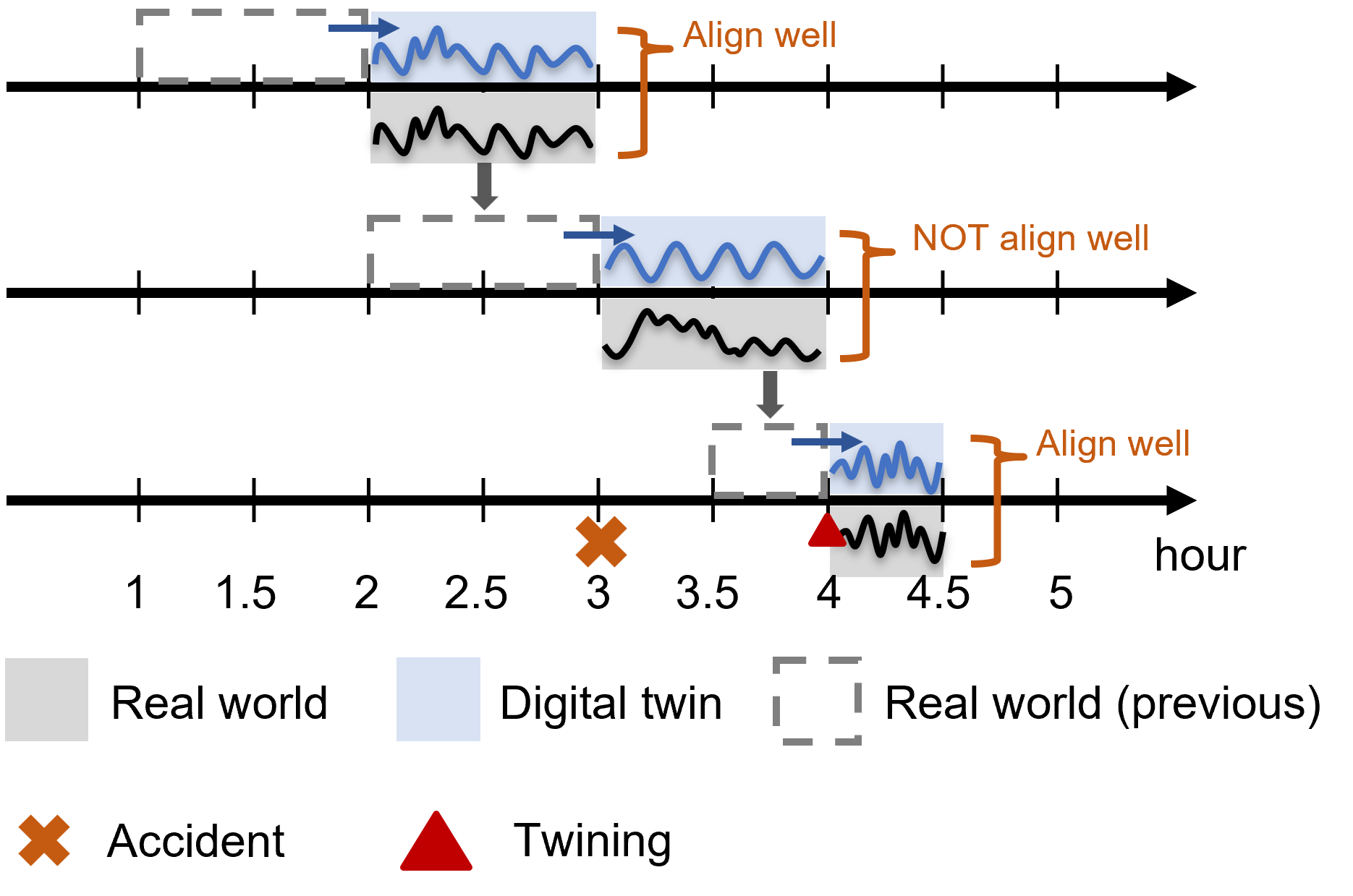}
    \caption{The long-short term twinning (LSTT) component workflow. Starting with the long-term twinning, LSTT activates the short-term twining when the simulated traffic states deviate from the observations. Once the deviation diminishes, LSTT switches back to the long-term twining.}
    \label{fig:LSTT component}
\end{figure}

In summary, long-term twining is cost-efficient but may incur a loss of fidelity, while the short-term one is more responsive to traffic disturbances and offers accurate replicas at the cost of higher computation overhead. To get the best of two worlds, we propose a long-short-term twining (LSTT) mechanism, a self-adaptive twining scheme that automatically switches between long and short-term periods.  Starting with long-term twinning, LSTT ensures high-quality safety risk simulation by switching to short-term twinning when misalignment with actual traffic conditions is detected, and reverting to long-term twinning when alignment improves. \Cref{fig:LSTT component} presents a schematic illustration of the LSTT workflow.

The key step in LSTT is the simulation fidelity assessment: whether the current simulated safety risk matches the ground truth, which triggers the long/short-term switch. Since the actual safety risk (SSM) in the real world is unobservable to the PTCs, a direct comparison is invalid. Hence, for every 30 minutes (consistent with the short-term period), LSTT compares the average simulated traffic volume and speed with the fusion data average. For the $k$-th 30-minute period, we define the volume deviation on edge $(i,j)$ as $\operatorname{SimDev}(i,j)\triangleq |\sum_{\tau=0}^{29}[\hat{q}^*_{30k+\tau}(i,j)-q^f_{30k+\tau}(i,j)]/\sum_{\tau=0}^{29}(q^f_{30k+\tau}(i,j))|$, which is essentially a MAPE of the 30-minute average of MTSS volume simulation with respect to that of fusion volume. The speed deviation can be defined accordingly. 

As summarized in the LSTT part in \Cref{fig:DT-workflow}, LSTT proceeds by selecting edges whose predicted SSM is above a certain threshold, which is 3 in our experiment, and deriving the $\operatorname{SimDev}$ for each edge. The rationale behind this selection is that MTSS is designed to replicate the driving risks. Therefore, edges with high SSM are of primary interest (referred as candidate edges) when inspecting the simulation fidelity: if there exists an edge $(i,j)$ from the set of candidate edges, such that the deviation is above the triggering threshold $\alpha$, i.e., $\operatorname{SimDev}(i,j)\geq \alpha$, then LSTT asserts that the current long-term period renders the simulation no longer align well with the actual traffic situation and the resulting SSM prediction does not reflect the actual driving risks. Consequently, the short-term twining is activated. Conversely, after switching the short-term twining, if $\operatorname{SimDev}$ for each candidate edge falls below $20\%$ for two consecutive short-term periods, LSTT will revert back to the long-term twining to save computational resources, as shown in \Cref{fig:LSTT component}.

\subsection{Risk-aware intervening service}


\label{subsec:camera-surveillance}

%
The risk-aware intervening service is tasked to monitor edges with significant mobility and safety risks via the multi-PTC system. From a system engineering point of view, the intervening service, taking in the SSM predicted from MTSS and the fused traffic states from STTE, outputs real-time PTC tilting actions that enable PTCs within the network to collaboratively monitor high-risk edges. In order to deliver a real-time cooperative camera-tilting control, we propose a risk-constrained correlated online learning (RiCCOL) method, and its workflow is presented in \Cref{fig:DT-DIMA workflow}. Before jumping into the algorithmic details of RiCCOL, we begin by formulating the risk-aware surveillance problem in mathematical terms.      

\paragraph{Problem Formulation} The PTCs aim to monitor high-risk edges. In this case, the quantity of interest is the mobility and safety risk on every edge. The mobility risk refers to the abrupt changes in traffic mobility patterns, characterized by the difference between two consecutive traffic states $\bm{s}_{t}-\bm{s}_{t-1}$. Its normalization $\bm{\delta}_t\triangleq (\bm{s}_{t}-\bm{s}_{t-1})/\|\bm{s}_{t}-\bm{s}_{t-1}\|_1\in \R^{|\mathcal{E}|}$ reflects the network-wide traffic fluctuation distribution and its $(i,j)$-entry $\bm{\delta}_t(i,j)$ denotes the percentage of the fluctuation on the corresponding edge over the overall fluctuation. The higher $\bm{\delta}_t(i,j)$ is, the more significant mobility risk the edge carries. In addition to the mobility risk, the safety risk is of great concern to the surveillance camera. Denote by $\bm{r}_t\in \R^{|\mathcal{E}|}$ the actual SSM distribution, defined similarly as in \Cref{Eq: SSM_Count} with the numerator and denominator replaced by the ground truth.

Given the PTCs' joint action $\bm{a}_t$, the percentage of captured mobility and safety risks of those monitored edges are given by  $f(\bm{a}_t; \bm{\delta}_t)\triangleq \langle \bm{a}_t, \bm{\delta}_t \rangle $ and  $g(\bm{a}_t; \bm{r}_t)\triangleq \langle \bm{a}_t, \bm{r}_t \rangle/\|\bm{r}_t\|_1$, respectively. Even though the actual mobility and safety risks are not fully observable to PTCs (e.g., edges out of sight), their estimates are available thanks to the driver-informed predictive service. Specifically, the traffic fluctuation distribution can be approximated using the fusion data $\hat{\bm{\delta}}_t\triangleq (\bm{s}^f_{t}-\bm{s}^f_{t-1})/\|\bm{s}^f_{t}-\bm{s}^f_{t-1}\|_1$, while the predicted SSM $\hat{\bm{r}}_t$ by MTSS approximates the true SSM. Informed by STTE and MTSS, the PTC controller is aware of the approximate risks across the network and tilts the camera accordingly.

Contrary to traditional mobility management strategies that focus on mobility risks, our risk-aware intervening service prioritizes safety risk management. Therefore, we formulate the PTC control problem as a constrained optimization in \Cref{eq:constrained-opt}, where the safety-risk constraint mandates PTCs to prioritize the surveillance on the edges with significant safety risks and capture at least $\beta$ percentage of SSM. When such a precondition is met, the multi-PTC system will dedicate the rest of the cameras to mobility-risk surveillance to maximize the traffic fluctuation capture. Note that the threshold $\beta$ is a hyperparameter to be configured by the TMC based on its needs. The higher $\beta$ is, the more cameras are devoted to monitoring edges with high SSM, and the less sensitive PTCs are to mobility risks.
\begin{equation}
\label{eq:constrained-opt}
    \max_{\bm{a}_{t}}   f(\bm{a}_t; \hat{\bm{\delta}}_t)\quad \text{s.t. }   g(\bm{a}_t; \hat{\bm{r}}_t) \geq \beta, t\in [T].
\end{equation}
 Unlike standard optimization where the objective and constraint functions are known in prior, the functions $f(\cdot; \hat{\bm{\delta}}_t)$ and $g(\cdot; \hat{\bm{r}}_t)$ are time-varying and are available only after the action is executed: $\hat{\bm{\delta}}_t$ relies on the fusion traffic state $s^f_t$ that is obtained after the PTC controller implements $\bm{a}_t$ and collects $\bm{s}_t^c$. The controller must learn from the previous feedback to determine the current action, and \Cref{eq:standard-opt} is referred to as online learning with constraints in the literature \cite{mannor06online-constraint}.

 Online learning, as a subfield of reinforcement learning \cite{tao22confluence, Tao_blackwell}, relaxes the Markovian state transition assumption as in standard RL formulation (i.e., Markov decision processes). The sequence of traffic fluctuations $\{\bm{\delta}_t\}$ are assumed arbitrarily generated without following any statistical rules, which captures the abrupt traffic disturbances caused by unexpected incidents. In addition to better modeling, online learning also presents a computationally lightweight solution, compared with RL approaches. As presented in the ensuing paragraph, the online learning control enjoys a plug-and-play operation without offline training or prior knowledge of the network topology or traffic patterns, albeit STTE and MTSS need to be configured to the network. In contrast, RL requires pre-training or fine-tuning when deployed in various traffic environments and can fail if there exists a distribution shift between training and deployment environments. 

 Besides RL, another competitor to our proposed online learning control is model predictive control (MPC) (also called rolling/receding horizon control) \cite{mesbah18stochasticMPC}, which optimizes the control policy over a lookahead horizon by predicting the system states in the future. When the system model is unknown,  advanced machine learning models or simulators (e.g., STTE and MTSS) play the role of predictor that forecasts the future states within the lookahead horizon. Even though this learning-based MPC has achieved encouraging success in many domains \cite{learning_MPC, tao23cola, hammer24col},  it is deemed overkill for the risk-aware intervening service for the following reasons. First, the PTC tilting actions do not affect the traffic state evolution, and there is no statistical correlation among states. Hence, optimizing the policy over a lookahead horizon is equivalent to optimizing over each step as in \Cref{eq:constrained-opt}. MPC's looking ahead into the future does not fit our problem formulation.  Second, since the action set is discrete, the MPC problem leads to a combinatorial optimization that depends on the underlying network, and the resulting computation complexity is of, at best, polynomial time. However, our proposed RiCCOL runs a distributed online learning control, and each PTC's computation complexity grows linearly with respect to its total number of actions (i.e., the number of edges); in the transportation network environment, one key intersection usually conjuncts 4 or 5 road segments. Therefore, our proposed online learning control is more scalable to extensive traffic networks.

\paragraph{Risk-constrained Correlated Online Learning}To facilitate our discussion of the algorithm design, we first treat the centralized control scenario, in which a single controller directs all PTCs within the network. The intuition behind the algorithm design is to transform the constrained optimization into a minimax problem of the corresponding Lagrangian, which then leads to a first-order primal-dual algorithm \cite{andrzej06nonlinear}. Towards this end, we first convert \Cref{eq:constrained-opt} to the standard form. Let $h(\bm{a}_t; \hat{\bm{r}}_t)\triangleq \beta -g(\bm{a}_t; \hat{\bm{r}}_t)$, and define the control policy $\pi\in\Pi$ as a Borel probability measure over the joint action space $\{0,1\}^{|\mathcal{E}|}$ , from which the action is sampled and executed. Since the joint action space is finite, the policy space is simply the probability simplex $\Pi\triangleq \Delta^{|\mathcal{E}|-1}$ Then, the constrained optimization in \Cref{eq:constrained-opt} is equivalent to 
\begin{equation}
\label{eq:standard-opt}
    \min_{\pi\in \Pi}   -\E_{a_t\sim \pi} f(\bm{a}_t; \hat{\bm{\delta}}_t)\quad \text{s.t. }   \E_{a_t\sim \pi }h(\bm{a}_t; \hat{\bm{r}}_t) \leq 0, t\in [T].
\end{equation}

Similarly to the Lagrangian multiplier method in mathematical programming, the online optimization \Cref{eq:standard-opt} can be translated to an unconstrained one through the Lagrangian. Let $\lambda$ be the non-negative Lagrangian multiplier associated with the constraint. The optimal solution to \Cref{eq:standard-opt} is characterized by the saddle point of the Lagrangian $\mathcal{L}(\pi,\lambda; \hat{\bm{\delta}}_t, \hat{\bm{r}}_t)$ defined below $$\min_{\pi\in \Pi}\max_{\lambda\in \R_{+}}\mathcal{L}(\pi,\lambda; \hat{\bm{\delta}}_t, \hat{\bm{r}}_t)\triangleq -f(\pi;\hat{\bm{\delta}}_t)+\lambda h(\pi;\hat{\bm{r}}_t).$$

A well-established method to solve the minimax problem above is the primal-dual gradient method \cite{andrzej06nonlinear}, which proceeds by applying gradient descent and ascent to the primal variable $\pi$ and the dual variable $\lambda$, respectively, as presented in \Cref{eq:primal-dual}.
\begin{subequations}
\label{eq:primal-dual}
\begin{equation}
\label{eq:primal}
    \pi_{t+1}= \operatorname{Proj}_{\Pi}[\pi_t-\gamma_t \nabla_\pi \mathcal{L}(\pi_t,\lambda_t;\hat{\bm{\delta}}_t,\hat{\bm{r}}_t)],
\end{equation}
\begin{equation}
\label{eq:dual}
        \lambda_{t+1}=\operatorname{Proj}_{R_{+}}[\lambda_t+\gamma_t \nabla_\lambda \mathcal{L}(\pi_t,\lambda_t;\hat{\bm{\delta}}_t,\hat{\bm{r}}_t)],
\end{equation}
\end{subequations}
where $\gamma_t$ denotes the step size in gradient descent-ascent. Taking a closer look at the projection in the primal update in \Cref{eq:primal}, we obtain 

\begin{align*}
 \operatorname{Proj}_{\Pi}[\pi_t-\gamma_t \nabla_\pi \mathcal{L}(\pi_t,\lambda_t;\hat{\bm{\delta}}_t,\hat{\bm{r}}_t)] =\argmin_{\pi\in \Pi}\left\{\gamma_t \langle \nabla_\pi \mathcal{L}(\pi_t,\lambda_t;\hat{\bm{\delta}}_t,\hat{\bm{r}}_t), \pi-\pi_t \rangle +\frac{1}{2}\|\pi-\pi_t\|_2^2\right\},
    \end{align*}
where $\|\cdot\|_2$ denotes the Euclidean distance, which fails to exploit the geometry of the probability simplex\cite{nemirovsky83md}. 

One remedy, as pioneered by \cite{nemirovsky83md}, is to replace the Euclidean distance with the entropy-based Bregman divergence. Given a negative entropy function $\varphi(\pi)=\sum_{a\in \mathcal{A}}\pi(a)\ln\pi(a)$, the Bregman divergence with respect to $\varphi$ is defined as $D_\varphi(y,x)\triangleq\varphi(y)-\varphi(x)-\langle \nabla \varphi(x), y-x \rangle$, and the resulting primal update becomes
$\pi_{t+1} = \argmin_{\pi\in \Pi}\{\gamma_t \langle \nabla_\pi \mathcal{L}(\pi_t,\lambda_t;\hat{\bm{\delta}}_t,\hat{\bm{r}}_t), \pi-\pi_t \rangle +D_\varphi(\pi,\pi_t)\}$. Thanks to the entropy function, the primal update admits a closed-form expression (which does not hold for the Euclidean distance):
\begin{align}
        \pi_{t+1}(a)=\frac{\pi_t(a)\exp\{-\gamma_t \nabla_\pi \mathcal{L}(\pi_t,\lambda_t;\hat{\bm{\delta}}_t,\hat{\bm{r}}_t)[a]\}}{\sum_{a'\in \mathcal{A}}\pi_t(a')\exp\{-\gamma_t \nabla_\pi \mathcal{L}(\pi_t,\lambda_t;\hat{\bm{\delta}}_t,\hat{\bm{r}}_t)[a']\}},
    \end{align}
 where $\nabla_\pi \mathcal{L}(\pi_t,\lambda_t;\hat{\bm{\delta}}_t,\hat{\bm{r}}_t)[a]\triangleq -f(a; \hat{\bm{\delta}}_t)+\lambda_t h(a;\hat{\bm{r}}_t)$ denotes the entry corresponding to the action $a$. Plugging the Lagrangian's gradient back into the primal update, we obtain
 \begin{equation}
 \label{eq:mw-primal}
     \pi_{t+1}(a)=\frac{\pi_t(a)\exp\{ \gamma_t[ f(a; \hat{\bm{\delta}}_t)-\lambda_t h(a;\hat{\bm{r}}_t)] \}}{\sum_{a'\in \mathcal{A}}\pi_t(a')\exp\{\gamma_t [f(a'; \hat{\bm{\delta}}_t)-\lambda_t h(a';\hat{\bm{r}}_t)] \}},
 \end{equation}
which coincides with the seminal multiplicative weights algorithm in online learning literature \cite{bianchi02adv-bandit}. Actions leading to higher exponential weights $\exp\{ \gamma_t[ f(a; \hat{\bm{\delta}}_t)-\lambda_t h(a;\hat{\bm{r}
}_t)] \}$, that is higher mobility-risk coverage and less safety-risk constraint violation, will be assigned more probability mass, i.e., greater $\pi_{t+1}(a)$, and chosen more frequently in the subsequent.   

Since the dual variable belongs to the real line, the projection admits a simple expression:
\begin{align}
\label{eq:dual-proj}
        \lambda_{t+1}=\max\{0, \lambda_t+\gamma_t \nabla_\lambda \mathcal{L}(\pi_t,\lambda_t;\hat{\bm{\delta}}_t,\hat{\bm{r}}_t)\}=\max\{0, \lambda_t+\gamma_t h(\pi_t;\hat{\bm{r}}_t)\}.
\end{align}
The key observation from the dual update is that once the violation of the constraint is observed, i.e., $h(\pi_t;\hat{\bm{r}}_t)>0$, $\lambda_{t+1}$ increases and more focus is placed on the safety-risk constraint violation in the exponential weights in \Cref{eq:mw-primal}. Consequently, PTCs are tilted to monitor the edges with high safety risks.

However, such a centralized control scales poorly when facing large-scale networks. To alleviate the curse of dimensionality in PTC control, we propose a distributed control algorithm based on the primal-dual updates in \Cref{eq:mw-primal} and \Cref{eq:dual-proj}, which enables parallel primal-dual updates at each PTC.  The key challenge of this distribution lies in the proper assignment of the exponential weights to each PTC's individual actions.  

Taking inspiration from correlated learning in game-theoretic learning literature \cite{shutian23erm}, we propose correlated exponential weight to assess the contribution of each PTC's action to network-wide mobility/safety-risk surveillance. Towards this end, we first revisit the saddle point of the Lagrangian from an individual PTC's perspective: $$\min_{\pi^i\in \Delta(\mathcal{A}^i)}\max_{\lambda^i\in \R_{+}}\mathcal{L}^i(\pi^i,\lambda^i; \hat{\bm{\delta}}_t, \hat{\bm{r}}_t)\triangleq -f(\pi^i, \bm{a}_t^{-i};\hat{\bm{\delta}}_t)+\lambda^i h(\pi^i, \bm{a}_t^{-i};\hat{\bm{r}}_t),$$ where the decision variables $\pi^i$ and $\lambda^i$ are all local variables with respect to the single PTC and the joint control actions of all other cameras, denoted by $\bm{a}_t^{-i}$, are fixed. 

Following the same argument above, fixing all others' actions, the exponential weight for the $i$-th PTC is given by $\exp\{ \gamma_t[ f(a^i, \bm{a}_t^{-i}; \hat{\bm{\delta}}_t)-\lambda^i_t h(a^i, \bm{a}_t^{-i};\hat{\bm{r}}_t)] \}$. Note that such individual weight correlates with others' actions, and hence, the primal update of each individual is statistically correlated with the rest of the PTCs, achieving intra-camera coordination. Plugging the correlated weight into the primal-dual update in \Cref{eq:mw-primal} and \Cref{eq:dual-proj}, we obtain the correlated primal-dual update for each PTC.
\begin{subequations}
\label{eq:riccol}
\begin{align}
    \label{eq:corr-primal}
     \pi^i_{t+1}(a)&=(1-\epsilon_t)\frac{\pi^i_t(a)\exp\{ \gamma_t[ f(a, \bm{a}_t^{-i}; \hat{\bm{\delta}}_t)-\lambda^i_t h(a, \bm{a}_t^{-i};\hat{\bm{r}}_t)] \}}{\sum_{a'\in \mathcal{A}^i}\pi^i_t(a')\exp\{\gamma_t [f(a', \bm{a}_t^{-i}; \hat{\bm{\delta}}_t)-\lambda^i_t h(a', \bm{a}_t^{-i};\hat{\bm{r}}_t)] \}}+\epsilon_t \frac{1}{|\mathcal{A}^i|},\\
     \label{eq:corr-dual}
        \lambda^i_{t+1}&=\max\{0, \lambda^i_t+\gamma_t h(a^i_t, \bm{a}_t^{-i};\hat{\bm{r}}_t)\}.
\end{align}
\end{subequations}
Of note is that \Cref{eq:corr-primal} admits an $\epsilon$-exploration term, which is a standard practice in online learning \cite{bianchi02adv-bandit}, encouraging exploration in the presence of unexpected traffic disturbances.

\subsection{Synchronization of Inner Components in DT-DIMA}
Due to distinct computation and simulation capacities, STTE, MTSS, and RiCCOL operate on asynchronous timescales. On the slowest timescale, MTSS is called once every $P^*$ minute, where $P^*=30$ or $60$ minutes, depending on whether the short-term twining is triggered. STTE operates on a faster timescale. It is called $K=P^*/\phi$ times within a $P^*$-minute interval, where $\phi$ denotes the aggregation level: each STTE prediction data point represents the average of $\phi$-minute traffic states. RiCCOL, as the multi-PTC controller, runs on a minute-by-minute basis, i.e., on the fastest timescale. The outputs of MTSS and STTE are synchronized according to the RiCCOL's timescale to ensure a smooth synergy of the three. We consider the following instance to unveil such synergy, where MTSS begins a new simulation period at time $t$ and $P^*=60$. 

One MTSS run simulates the traffic volume and speed at each time step in future $P^*$ minutes, and hence, the corresponding output of MTSS includes $\{\hat{\bm{s}}^*_\tau\}_{\tau\in [t, t+P^*)}$, and the traffic states could be volume and speed $\hat{\bm{s}}^*=\hat{\bm{q}}^*/ \hat{\bm{v}}^*$. Another important output by MTSS is the SSM introduced in \Cref{subsubsec: MTSS}. The SSM $\hat{\bm{r}}_k$ for the $k$-th $ d_t$ interval is calculated every $ d_t$ minutes and then expanded to the $P^*$-intervel on a minute basis: $\hat{\bm{r}}_{\tau}\triangleq \hat{\bm{r}}_{k}$, if $\tau\in [k d_t, (k+1) d_t)$. 

STTE works in a rolling manner: its one-time output includes $K$ predictions corresponding to future traffic states in $P^*$ minutes since each prediction represents a $\phi$-minute average. Similar to the expansion of SSM, we break down the aggregated data into a minute level: for the $k$-th prediction data point $\hat{\bm{s}}_k$ by STTE at time $t$, let $\hat{\bm{s}}_\tau \triangleq \hat{\bm{s}}_k$, $\tau\in [k\phi, (k+1)\phi)$. When STTE is called again after $\phi$ minutes (at $t+\phi$), it takes in the most recent fusion states for the past $P^*$ minutes, which is aggregated on the $\phi$-minute basis to ensure the input and output are of the same dimension, and then gives the predictions for the interval $[t+\phi, t+\phi+P^*)$.  Of important remark is that the first batch of $K$ predictions is fed to MTSS for updates of future stages (see \Cref{subsubsec: MTSS}).


\section{Experiment}
\subsection{Dataset summary}
\begin{figure}[t!]
    \centering
    \includegraphics[width=0.98\columnwidth]{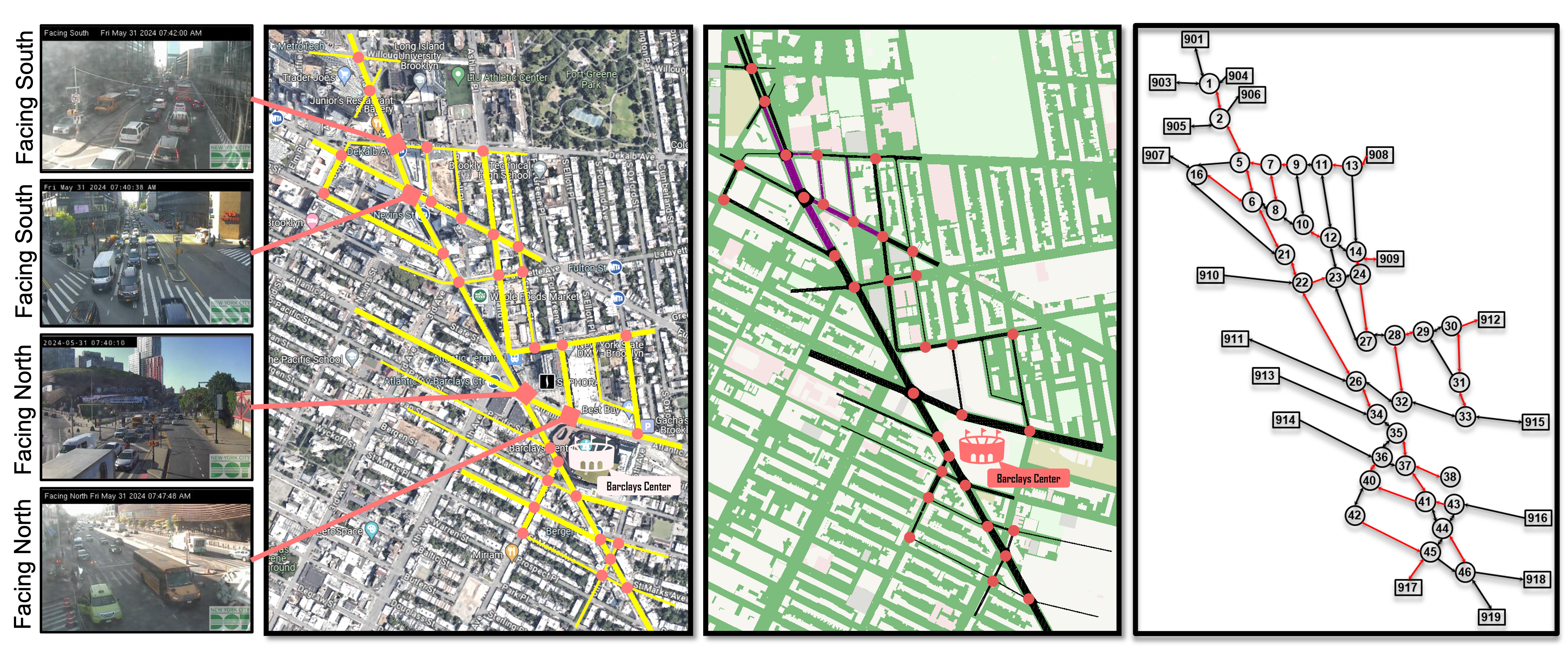}
    \caption{The experimental networks in DT-DIMA. Left to right: Example frames from current PTCs, Open Street Map, SUMO, Abstract Network, the red edges indicate the road segments covered by the base model in \Cref{subsubsec: basemodel}.}
    \label{fig:exp-network}
\end{figure}

\paragraph{Road network configuration} For our study, we have carefully constructed a simulation environment in Brooklyn, New York, replicating the dynamics of the Flatbush Avenue corridor, extending from Willoughby Street to Grand Army Plaza. This corridor includes complex urban traffic challenges, including proximity to a large sports arena, the Barclays Center, making it a prime subject for our simulation. The simulation setup and its real-world counterpart are illustrated in \Cref{fig:exp-network}. The modeled network comprises 117 road segments and 38 intersections. We assume the presence of one PTC at each intersection, resulting in a total of 38 PTCs.

\paragraph{Calibration of the simulation in MTSS} 
To mirror the real world in DT using MTSS component and replicate accurate driving behaviors, we calibrate the simulation in MTSS component using a blend of real-world data sourced from various data points. Key parameters such as roadway geometry, lane usage, edge capacities, speed limits, and turning connectivities are updated to reflect realistic conditions as of 2018. In addition to network topology, essential road network details like signal timings and bus stop locations are incorporated into our simulation network. We also include surrogate safety metrics to enhance our understanding of traffic dynamics and safety scenarios.

The calibration is conducted for operational measures, including edge volumes, travel times, and safety measures, like conflicts identified by time-to-collision metrics. We employ the Simultaneous Perturbation Stochastic Approximation (SPSA) method to ensure our simulation accurately mirrors the traffic conditions and behaviors observed on the Flatbush corridor \cite{sha2020applying}. During real-time operation of DT-DIMA, we run 10 iterations of simulations with different random seeds for each MTSS call. The outputs of MTSS are then averaged with calculated standard deviation.

\paragraph{Dataset preparation}
The training dataset for the STTE component includes edge-level traffic volume and traffic speed information at 1-minute intervals. The dataset is generated by SUMO, which comprises data representing 200 days of standard traffic conditions. The data for each day covers 24 hours with one-second intervals, providing a high-resolution view of daily traffic patterns and variations. We generate traffic data that includes average volume and average speed, recorded on a minute-by-minute basis. Hence, each time step represents one minute, and the horizon length is $T=1440$.  This granularity allows for a nuanced understanding of traffic flow and speed patterns across different segments of the road network. To test the DT-DIMA's robustness and responsiveness to different traffic conditions, we introduce 38 traffic incident scenarios involving random lane closures ranging from one to five hours. 

The testing data is generated using the calibrated simulation environment. Each testing scenario uses a new random seed to ensure the generated data are not used in STTE training. Unless otherwise specified, all experiment results are obtained from 20 repeated experiments with different random seeds. 

\subsection{Experiment setup}\label{subsec:experiment_setup}
\paragraph{Configuration of STTE and RiCCOL} The STTE component uses batch learning with a batch size of 20, dividing the training data into 60\% training, 20\% validation, and 20\% testing, normalized within a [-1,1] range. It operates on a Windows 11 desktop with an AMD Ryzen 9 7900X, 64 GB RAM, and NVIDIA RTX 3090Ti, utilizing the Pytorch library. Key settings include $h=8$ attention heads, a learning rate $lr=0.001$, and a model dimension $d_{model}=64$. Traffic data is collected every 5 minutes and analyzed over one-hour periods. As such, each prediction instance denotes the average traffic state forecast for a 5-minute interval ($\phi=5$), the controller uses the one prediction instance repeatedly five times within the interval until a new prediction is generated for the next 5-minute interval. One prediction instance's batch size is $K=12$, so the input/output sequence spans one hour. Consequently, the first hour of the day is the warm-up period, where each PTC uniformly picks one edge to monitor at each time step. The starting time is 1 am, and the proposed RiCCOL algorithm takes over afterward.

The hyper-parameters involved in the RiCCOL algorithm \Cref{eq:riccol} are as below. The safety-risk coverage threshold is $\beta=0.7$. The primal-dual update admits a constant step size $\gamma_t=1$. The exploration rate is $\epsilon_t=0.3$. The triggering threshold in LSTT is $\alpha=20\%$.
\paragraph{Configuration of simulated traffic incidents} To evaluate the robustness, resilience, and accuracy of the proposed DT-DIMA system, we introduced three types of traffic situations. These scenarios were designed to test the system's performance in responding to various simulated traffic incidents, providing a comprehensive assessment of its capabilities.

\begin{itemize}
    \item \textbf{Normal}: A simulated day with normal traffic situations, including recurrent traffic patterns such as morning and evening peak.
    \item \textbf{ACC}: A simulated day with one unexpected traffic accident happened on Edge 26-22 lasting one hour from 1:15 PM to 2:15 PM. The accident closed two out of three lanes.
    \item \textbf{WZ}: A simulated day with a pre-planned work zone with two-lane closure happened on Edge 22-21 lasting two hours from 6:00 AM to 8:00 AM. The work zone closed two out of three lanes.
\end{itemize}

\subsection{Evaluation for DT-DIMA performance}
\subsubsection{Base model: current implementation of PTCs in real world}\label{subsubsec: basemodel}
To evaluate the performance of the proposed DT-DIMA system, we compare it with a base model designed based on the current implementation of PTCs in the real transportation network. According to \cite{nyctmcRealTime}, the current PTCs adopt a fixed tilting strategy and are manually controlled by the transportation operation center, primarily for monitoring major traffic corridors. Typically, these PTCs remain in the same direction unless adjusted by transportation managers for specific monitoring purposes. Consequently, the base model replicates this real-world implementation by using a fixed tilting strategy. The leftmost of \Cref{fig:exp-network} shows an example of some current PTCs in our experimental network.

In this study, we select the tilting directions of PTCs and maintain them throughout the entire day. For PTCs on Flatbush Ave (a major corridor), we constrain the tilting directions to be along Flatbush Ave. For PTCs on non-major roads, we randomly choose one direction from all possible directions (all road segments connected to the intersection). The randomly selected fixed tilting directions are illustrated in the rightmost of \Cref{fig:exp-network}, covering 58 (shown in red) out of 117 edges while the road segments along the Flatbush Ave are mostly covered by PTCs. 

It is important to note that the base model, without STTE and MTSS, does not offer the driver-informed predictive service and risk-aware intervening service featured in our proposed DT-DIMA system. Consequently, it only provides observed traffic mobility information. Therefore, the primary focus of the comparison between the base model and our proposed DT-DIMA is on how the current network-wide traffic states can be captured by both models.

\subsubsection{Evaluation of mobility risk}\label{subsubsec: mobilityrisk}

To evaluate the mobility risk identified by the proposed DT-DIMA system, we examine both the network-level fusion of current traffic states and the prediction of future traffic states. The STTE component within the driver-informed predictive service is responsible for fusing and forecasting traffic states, effectively capturing traffic dynamics and abrupt changes across various scenarios.


We use the mean absolute percentage error (MAPE) to evaluate the performance of STTE in estimating and predicting traffic volumes compared to actual data. Using the notations defined in \Cref{tab:notations}, the metric is defined in \Cref{eq:fore-fu-mae}, where $k$ represents the hour index and $\bm{s}_k$ and $\hat{\bm{s}}_k$ denote the average of states and predicted states within the $k$-th hour, respectively.  The hourly forecasting MAPE is denoted as $ForecastingMAPE$, and the hourly fusion MAPE is denoted as $FusionMAPE$:

\begin{equation}
\label{eq:fore-fu-mae}
    \begin{aligned}
        &ForecastingMAPE_k ={\|\bm{\hat{s}}_k-\bm{s}_k\|_1}/\| 
        \bm{s}_k\|_1,\\
        &FusionMAPE_k = {\|\bm{s}^f_k-\bm{s}_k\|_1}/\| \bm{s}_k\|_1.
    \end{aligned}
\end{equation}


We evaluate the STTE outputs in three scenarios: Normal, ACC, and WZ. The settings for these scenarios are described in \Cref{subsec:experiment_setup}. We compare the $FusionMAPE$ of our model with the base model as mentioned in \Cref{subsubsec: basemodel} across these scenarios. \Cref{tab:stte-fusionmape} shows the $FusionMAPE$ of traffic volume from our model and base model for each hour of the day. Lower MAPE values indicate better extraction of mobility risk information. We remark that the base model employs deterministic tilting actions, and hence, \Cref{tab:stte-fusionmape} does not report the standard deviations of the base's MAPE results.

Our findings indicate that our model's $FusionMAPE$ outperforms the base model across three scenarios, evaluated each hour of the day. The $FusionMAPE$ for our model ranges between $8.40\%$ and $15.11\%$, while the base model reports $FusionMAPE$ values between $40.63\%$ and $43.94\%$. This disparity is due to the base model's reliance on fixed tilting directions for PTCs, which capture traffic mobility from single directions and miss information from other directions. In contrast, the STTE component in our model works closely with PTCs to provide a network-wide estimation of traffic states, effectively complementing the uncovered directions in real time. The low $FusionMAPE$ values indicate that observations controlled by the RiCCOL component captured road segments with significant traffic fluctuations, and the fusion of the network-wide traffic states can reflect realistic traffic flow conditions in the physical world.

\Cref{tab:stte-predmape} indicates network-level traffic volume forecast provided by the STTE component, which is evaluated using $ForecastingMAPE$. Comparing the three scenarios, $ForecastingMAPE$ has seen similar values across all three scenarios over the entire day, ranging from $15.82\%$ to $23.94\%$. The WZ and ACC cases exhibit similar MAPEs as the Normal, ranging $16.95\%-21.40\%$ in 6-8 AM and $21.51\%-22.74\%$ in 1-2 PM, respectively. These results demonstrate that the STTE component effectively captures mobility risk under various conditions. The low $ForecastingMAPE$ highlights the STTE component's success in adopting the fused traffic states, learning their temporal dependencies while considering spatial correlations within the constraints of the road network topology, and providing accurate predictions.

\begin{table}[htp]
    \centering
    \begin{tabular}{clccccc}
    \toprule
       MAPE($\%$)     &        & 1 AM  & 2 AM & 3 AM & 4 AM & 5 AM\\
    \midrule
      \textbf{ \multirow{3}{*}{\rotatebox[origin=c]{90}{Base}}} &Normal   &  $42.11$ & $43.94$ & $43.71$ & $43.22$ & $41.69$\\
    &ACC            &  $42.11$ & $43.94$ & $43.71$ & $43.22$ & $41.69$\\
    &WZ           &  $42.11$ & $43.94$ & $43.71$ & $43.22$ & $41.69$ \\
    \midrule
       \textbf{ \multirow{3}{*}{\rotatebox[origin=c]{90}{STTE}}}   &Normal& $ {15.10\pm0.60}$ & $ {12.75\pm0.50}$ & $ {12.43\pm0.22}$ & $ {12.7\pm0.60}$ & $ {13.26\pm0.29}$\\
    &ACC           & $ {15.11\pm 0.60}$ &$ {12.78\pm 0.55}$ & $ {12.41\pm 0.20}$& $ {12.75\pm 0.60}$& $ {13.25\pm 0.30}$\\
    &WZ         & $ {13.39\pm0.54}$ & $ {12.14\pm0.73}$ & $ {11.86\pm0.52}$ & $ {12.21\pm0.25}$ & $ {12.39\pm0.39}$\\
    \toprule
        &        & 6 AM  & 7 AM & 8 AM & 9 AM & 10 AM\\
    \midrule
      \textbf{ \multirow{3}{*}{\rotatebox[origin=c]{90}{Base}}} &Normal     & $40.48$ & $40.80$ & $41.17$ & $42.09$&  $42.27$\\
    &ACC              & $40.48$ & $40.80$ & $41.17$ & $42.09$& $42.27$\\
    &WZ             & $40.77$ & $42.13$ & $41.35$ & $42.22$& $42.79$ \\
    \midrule
       \textbf{ \multirow{3}{*}{\rotatebox[origin=c]{90}{STTE}}}   &Normal& $ {13.12\pm0.38}$  & $ {12.72\pm0.31}$ & $ {11.08\pm1.23}$ & $ {8.98\pm0.96}$ & $ {9.63\pm0.85}$\\
    &ACC              & $ {13.13\pm0.39}$ & $ {12.75\pm0.26}$ & $ {10.58\pm 1.14}$ & $ {8.60\pm 1.07}$& $ {9.40\pm 1.07}$\\
    &WZ         & $ {11.11\pm0.36}$  & $ {10.14\pm0.63}$ & $ {9.40\pm0.78}$ & $ {8.40\pm0.26}$ & $ {9.80\pm0.23}$\\ 
    \toprule
        &        & 11 AM  & 12 PM & 1 PM & 2 PM & 3 PM\\
    \midrule
      \textbf{ \multirow{3}{*}{\rotatebox[origin=c]{90}{Base}}} &Normal   &  $41.65$  & $41.62$ & $41.38$ &$41.40$ & $40.93$\\
    &ACC            & $41.89$  & $42.05$ & $42.69$ &$42.22$ & $41.58$\\
    &WZ             & $41.89$ & $42.05$ & $42.69$ & $42.22$& $41.58$ \\
    \midrule
       \textbf{ \multirow{3}{*}{\rotatebox[origin=c]{90}{STTE}}}   &Normal& $ {10.54\pm0.86}$  & $ {10.77\pm0.17}$ & $ {10.66\pm0.74}$ & $ {9.70\pm0.51}$ & $ {10.56\pm0.68}$\\
    &ACC            & $ {10.36\pm0.83}$ & $ {10.55\pm 0.48}$ & $ {11.40\pm0.30}$ & $ {10.95\pm0.68}$ & $ {10.68\pm0.71}$\\
    &WZ         & $ {9.91\pm0.46}$  & $ {10.20\pm0.41}$ & $ {10.13\pm0.46}$ & $ {9.20\pm0.45}$ & $ {9.70\pm0.44}$\\ 
    \toprule
        &        & 4 PM  & 5 PM & 6 PM & 7 PM & 8 PM\\
    \midrule
      \textbf{ \multirow{3}{*}{\rotatebox[origin=c]{90}{Base}}} &Normal   &  $40.69$  & $41.28$ & $41.70$ & $42.06$ & $42.46$\\
    &ACC            & $40.63$  & $40.99$ & $41.35$ & $41.73$ & $42.54$\\
    &WZ            & $41.59$ & $40.94$ & $41.75$ & $42.09$& $42.90$  \\
    \midrule
       \textbf{ \multirow{3}{*}{\rotatebox[origin=c]{90}{STTE}}}   &Normal& $ {9.77\pm0.56}$  & $ {8.40\pm0.09}$ & $ {8.58\pm0.56}$ & $ {9.22\pm0.25}$ & $ {9.80\pm0.54}$\\
    &ACC           & $ {9.15\pm1.03}$ & $ {8.42\pm0.53}$ & $ {8.94\pm0.28}$ & $ {8.80\pm0.29}$&$ {9.14\pm0.35}$ \\
    &WZ         & $ {9.87\pm0.4}$ & $ {8.53\pm0.29}$ & $ {9.05\pm0.21}$ & $ {9.19\pm0.19}$ & $ {9.46\pm0.23}$\\ 
    \toprule
        &        & 9 PM  & 10 PM & 11 PM &  & \\
    \midrule
      \textbf{ \multirow{3}{*}{\rotatebox[origin=c]{90}{Base}}} &Normal   &$41.20$  & $42.29$ & $43.49$&  & \\
    &ACC            &$41.62$  & $41.71$ & $42.68$ &  & \\
    &WZ           &$41.81$  & $42.82$ & $43.46$ &  & \\
    \midrule
       \textbf{ \multirow{3}{*}{\rotatebox[origin=c]{90}{STTE}}}   &Normal&$ {9.82\pm0.17}$  & $ {9.26\pm0.13}$ & $ {10.38\pm0.49}$ &  & \\
    &ACC          & $ {9.30\pm 0.39}$ & $ {9.14\pm 0.37}$ &$ {9.85\pm 0.51}$  &  &\\
    &WZ         &$ {9.42\pm0.32}$ & $ {9.68\pm0.43}$ & $ {10.28\pm0.43}$ &  & \\ 
    \bottomrule
    \end{tabular}
    \caption{The MAPE results of network-level traffic volume in both base model and STTE. The MAPEs of the base model are represented as \textbf{Base}, and the MAPEs of STTE are represented as \textbf{STTE}. The MAPE results are presented every hour across three cases: Normal, ACC, and WZ. ACC is located on Edge 26-22, between 1:15 PM and 2:15 PM. WZ is located on Edge 22-21, between 6:00 AM and 8:00 AM.}
    \label{tab:stte-fusionmape}
\end{table}

\begin{table}[]
    \centering
    \begin{tabular}{clccccc}
    \toprule
       & MAPE($\%$)            & 1 AM  & 2 AM & 3 AM & 4 AM & 5AM\\
    \midrule
      \textbf{ \multirow{3}{*}{\rotatebox[origin=c]{90}{}}} &Normal   &  $23.17\pm0.54$ & $22.95\pm0.87$ & $23.44\pm0.67$ & $23.82\pm 0.93$ & $23.55\pm0.54$\\
    &ACC            & $23.19\pm0.54$ & $22.95\pm 0.88$ & $23.47\pm 0.67$ & $23.84\pm 0.94$& $23.56\pm 0.52$ \\
    &WZ           & $23.96\pm0.41$ & $22.16\pm1.27$ & $22.36\pm0.69$ & $23.03\pm0.99$ & $23.94\pm0.40$ \\
    \toprule
        &        & 6 AM  & 7 AM & 8 AM & 9 AM & 10 AM\\
    \midrule
      \textbf{ \multirow{3}{*}{\rotatebox[origin=c]{90}{}}} &Normal   &  $21.88\pm0.82$  & $23.01\pm0.19$ & $20.03\pm1.78$ & $16.76\pm1.16$ & $17.13\pm0.96$\\
    &ACC            & $21.90\pm 0.83$  & $23.09\pm0.30$ & $19.42\pm1.99$ &$16.26\pm1.28$ & $17.07\pm1.02$\\
    &WZ           & $21.40\pm0.82$  & $16.95\pm0.81$ & $18.11\pm1.20$ & $15.19\pm0.41$ & $17.57\pm0.50$ \\
    \toprule
        &        & 11 AM  & 12 PM & 1 PM & 2 PM & 3 PM\\
    \midrule
      \textbf{ \multirow{3}{*}{\rotatebox[origin=c]{90}{}}} &Normal   &  $20.25\pm1.16$  & $20.7\pm 0.75$ & $21.26\pm1.35$ & $19.64\pm1.11$ & $20.89\pm0.11$\\
    &ACC            & $20.14\pm1.12$  & $20.5\pm1.04$ & $21.51\pm0.53$ & $22.74\pm0.79$ & $21.15\pm0.86$\\
    &WZ           & $19.26\pm0.73$  & $19.37\pm0.93$ & $19.75\pm0.67$ & $18.67\pm0.68$ & $19.35\pm0.59$ \\
    \toprule
        &        & 4 PM  & 5 PM & 6 PM & 7 PM & 8 PM\\
    \midrule
      \textbf{ \multirow{3}{*}{\rotatebox[origin=c]{90}{}}} &Normal   &  $18.99\pm0.11$  & $15.90\pm 0.99$ & $17.03\pm0.22$ & $18.49\pm0.40$ & $19.36\pm1.23$\\
    &ACC            & $17.75\pm0.14$  & $15.82\pm0.05$ & $16.61\pm0.39$ & $17.56\pm0.43$ & $18.39\pm1.01$\\
    &WZ           & $18.68\pm0.56$  & $16.60\pm0.25$ & $16.66\pm0.27$ & $17.63\pm0.29$ & $18.15\pm0.50$ \\
    \toprule
        &        & 9 PM  & 10 PM & 11 PM &  & \\
    \midrule
      \textbf{ \multirow{3}{*}{\rotatebox[origin=c]{90}{}}} &Normal   &$19.34\pm0.66$  & $17.21\pm0.52$ & $20.32\pm0.55$&  & \\
    &ACC            &$18.69\pm 0.43$  & $16.81\pm 0.48$ & $18.85\pm 0.32$ &  & \\
    &WZ           &$18.83\pm0.76$  & $17.02\pm0.43$ & $19.64\pm0.49$ &  & \\
    \bottomrule
    \end{tabular}
    \caption{The MAPE results of the forecast of traffic volume in the STTE component. The MAPE results are presented every hour across three cases: Normal, ACC, and WZ. ACC is located on Edge 26-22, between 1:15 PM and 2:15 PM. WZ is located on Edge 22-21, between 6:00 AM and 8:00 AM.}
    \label{tab:stte-predmape}
\end{table}

\begin{figure}
    \centering
    \includegraphics[width=1\textwidth]{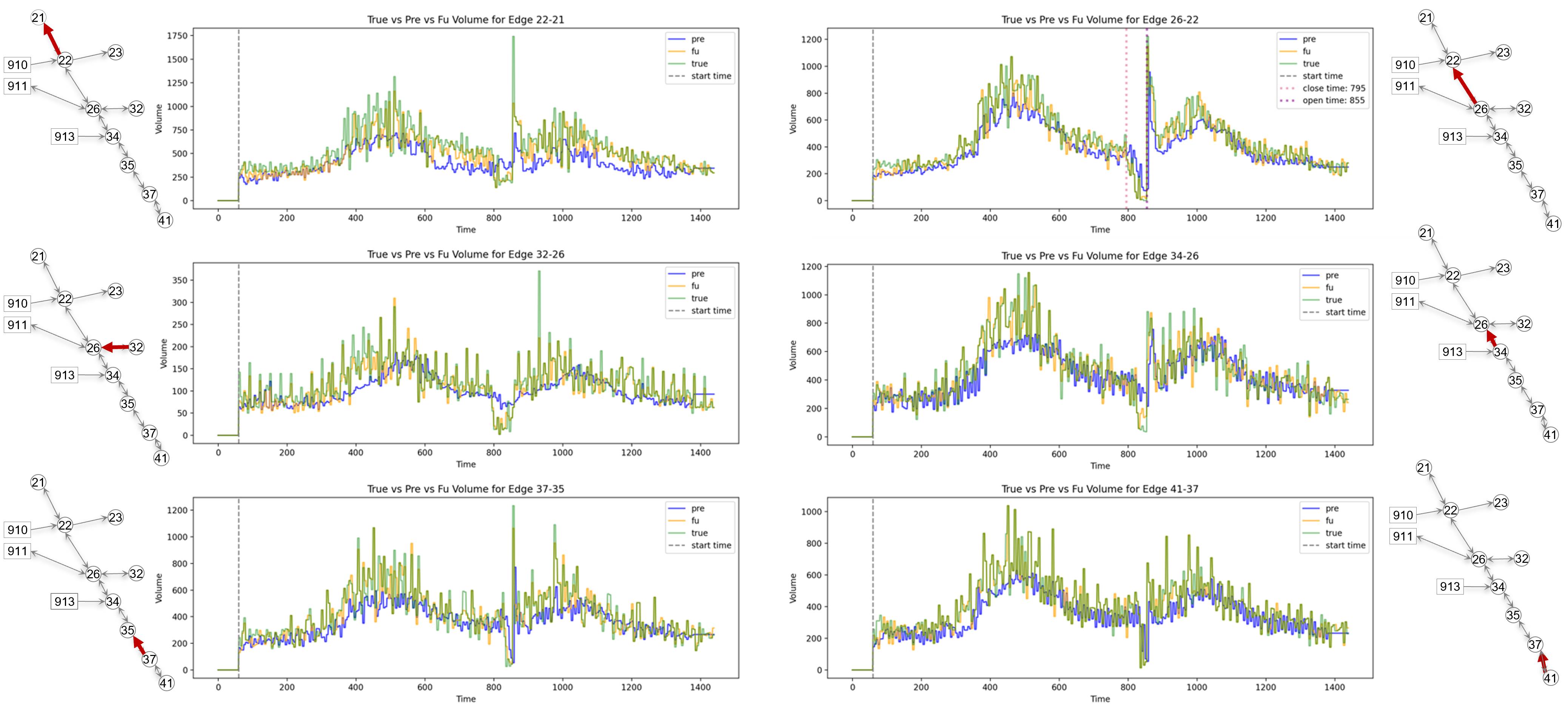}
    \caption{Selected edge traffic distribution from ACC. ACC is located on Edge 26-22 (upper right), between 1:15 PM and 2:15 PM.}
    \label{fig:acc67-stte}
\end{figure}

The visualization in \Cref{fig:acc67-stte} highlights the edge where the ACC occurred (Edge 26-22) and its neighboring edges. The green line represents the ground truth traffic volume, the orange line represents the fused traffic volume, and the blue line indicates the predicted traffic volume. The two red vertical dashed lines in the upper right of \Cref{fig:acc67-stte} mark the start and end times of the ACC. Both the fusion and prediction lines from the STTE component successfully capture the abrupt changes caused by the ACC, as well as the propagation of disruptions and queues to neighboring edges. This confirms that the STTE component can quickly respond to traffic disturbances and provide accurate estimations under various traffic situations.

\begin{figure}
    \centering
    \includegraphics[width=1\textwidth]{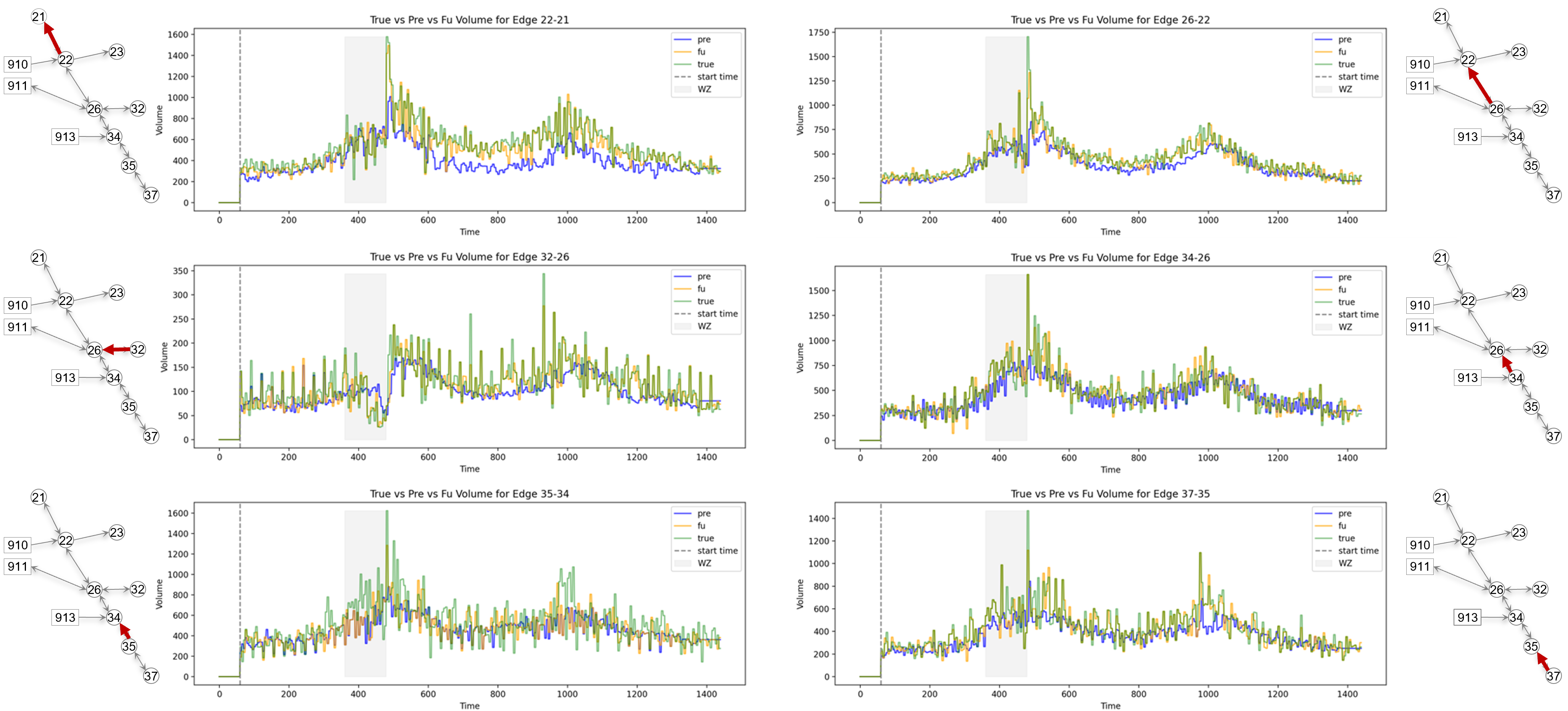}
    \caption{Selected edge traffic distribution from WZ. WZ is located on Edge 22-21 (upper left), between 6:00 AM and 8:00 AM.}
    \label{fig:wz168-stte}
\end{figure}

Similarly, \Cref{fig:wz168-stte} shows the edge where the work zone (WZ) was located (Edge 22-21) and its neighboring edges. Due to a two-hour, two-lane closure work zone between 6 AM and 8 AM, traffic volume during peak hours significantly decreased with an immediate rebound right after removing the WZ. It is evident in \Cref{fig:wz168-stte} that the fusion line matches well with the true data across all selected edges, while the prediction line has a relatively higher bias on WZ Edge 22-21 but matches well with the true data along other neighboring edges.

\subsubsection{Evaluation of safety risk}

This section evaluates the predicted safety risks provided by the MTSS component, represented using SSMs as detailed in \Cref{subsubsec: MTSS}. Higher SSM values along a road segment indicate greater safety risks. As illustrated in \Cref{fig:DT-workflow}, the MTSS component receives real-time fused traffic states from the STTE component and observed traffic incident information from the multi-PTC sensing service. The MTSS component uses this input to conduct real-time simulations replicating the physical world and predicting network-level SSMs.

To assess the MTSS component's effectiveness in predicting network-level SSMs, we consider its performance at two levels: 1) the forecasting level and 2) the reconstruction level. At the forecasting level, we evaluate the real-time prediction performance of the MTSS component. During the operation of the DT-DIMA system, acknowledging the latency exists as the simulation within the MTSS component is called every $P^*$ minutes, we compare each predicted SSM with true data to determine if the MTSS can accurately predict the safety risks. This evaluation tests both the predictability of SSMs by the MTSS component and the accuracy of the information provided by other parts of the DT-DIMA system.

At the reconstruction level, we evaluate the MTSS component's capability to replicate the physical world after receiving all traffic information for the entire day. Here, the MTSS component uses the exact settings experienced in the physical world to generate SSMs in a single process. We denote by $\bm{r}^f_t$ the reconstructed SSM, defined similarly as in \Cref{Eq: SSM_Count} with the numerator and denominator replaced by the reconstructed data. This evaluation tests the simulation capability of the MTSS component, ensuring it can accurately mirror the real-world conditions based on received traffic data. Towards this end, we employ the mean absolute percentage error to evaluate both the forecasting and reconstruction of SSMs in \Cref{eq:fore-fu-mape}. Denote by $\bm{r}_k$, $\hat{\bm{r}}_k$, and $\bm{r}^f_k$ the average of actual, forecasted, and reconstructed SSM within $k$-th hour, then the MAPEs are defined as  
\begin{equation}
\label{eq:fore-fu-mape}
    \begin{aligned}
        &ForecastingMAPE_k ={\|\bm{\hat{r}}_k-\bm{r}_k\|_1}/{\|\bm{r}_k\|_1},\\
        &ReconstructionMAPE_k = {\|\bm{r}^f_k-\bm{r}_k\|_1}/{\|\bm{r}_k\|_1}.
    \end{aligned}
\end{equation}

\begin{table}[htp]
    \centering
    \small
    
    \begin{tabular}{clccccc}
    \toprule
       (MAPE $\%$)     &        & 1 AM  & 2 AM & 3 AM & 4 AM & 5 AM\\
    \midrule
      \textbf{ \multirow{3}{*}{\rotatebox[origin=c]{90}{\tiny Forecast}}} &Normal   &  $1.02\pm 0.11$ & $1.13\pm 0.21$ & $1.03\pm 0.13$ & $1.42\pm 0.22$ & $2.19\pm 0.10$\\
    &ACC            & $1.01\pm0.11$ & $1.10\pm 0.18$ & $1.02\pm 0.14$ & $1.37\pm 0.15$& $2.23\pm 0.12$ \\
    &WZ           & $1.00\pm0.08$ & $1.15\pm 0.10$ & $0.97\pm 0.13$ & $1.42\pm 0.12$ & $2.18\pm 0.12$ \\
    \midrule
       \textbf{ \multirow{3}{*}{\rotatebox[origin=c]{90}{\tiny Reconstruct}}}   &Normal& $0.98\pm 0.03$ & $1.07\pm 0.08$ & $1.10\pm 0.03$ & $1.26\pm 0.07$ & $2.12\pm 0.14$\\
    &ACC          & $0.97\pm 0.03$ & $1.08\pm 0.08$ & $1.11\pm 0.03$ & $1.22\pm 0.08$& $2.13\pm 0.13$\\
    &WZ         & $1.01\pm 0.07$ & $1.12\pm 0.11$ & $1.11\pm 0.09$ & $1.21\pm 0.06$ & $2.10\pm 0.10$\\
    \toprule
        &        & 6 AM  & 7 AM & 8 AM & 9 AM & 10 AM\\
    \midrule
      \textbf{ \multirow{3}{*}{\rotatebox[origin=c]{90}{\tiny Forecast}}} &Normal   &  $6.30\pm 0.34$  & $8.87\pm 0.33$ & $8.77\pm 0.69$ & $5.06\pm 0.33$ & $2.68\pm 0.19$\\
    &ACC            & $6.23\pm 0.37$  & $8.58\pm0.70$ & $8.92\pm0.52$ & $5.02\pm 0.30$ & $2.74\pm 0.14$\\
    &WZ           & $ {6.98\pm 0.66}$  & $ {7.77\pm 0.48}$ & $ {12.97\pm 1.44}$ & $5.52\pm 0.57$ & $3.18\pm 0.24$ \\
    \midrule
       \textbf{ \multirow{3}{*}{\rotatebox[origin=c]{90}{\tiny Reconstruct}}}   &Normal& $3.88\pm 0.46$ & $4.92\pm 0.42$ & $5.45\pm 0.33$ & $3.66\pm 0.24$ & $2.72\pm 0.21$\\
    &ACC              & $3.95\pm0.37$ & $4.82\pm0.45$ & $5.44\pm 0.31$ & $3.61\pm 0.25$& $2.72\pm 0.22$\\
    &WZ         & $ {4.14\pm 0.28}$  & $ {5.89\pm 0.92}$ & $ {8.27\pm 1.61}$ & $4.53\pm 0.26$ & $2.60\pm 0.20$\\ 
    \toprule
        &        & 11 AM  & 12 PM & 1 PM & 2 PM & 3 PM\\
    \midrule
      \textbf{ \multirow{3}{*}{\rotatebox[origin=c]{90}{\tiny Forecast}}} &Normal   & $2.55\pm 0.16$  & $1.90\pm 0.11$ &  $2.12\pm 0.19$ & $2.34\pm 0.16$ & $4.22\pm 0.29$\\
    &ACC            & $2.57\pm 0.15$  & $1.95\pm0.15$ & $ {2.68\pm 0.17}$ &$ {5.10\pm0.11}$ & $3.86\pm 0.32$\\
    &WZ           & $2.55\pm 0.19$  & $1.80\pm0.05$ & $2.14\pm 0.06$ & $2.34\pm 0.37$ & $3.74\pm 0.22$ \\
    \midrule
       \textbf{ \multirow{3}{*}{\rotatebox[origin=c]{90}{\tiny Reconstruct}}}   &Normal& $2.74\pm 0.24$  & $1.97\pm 0.09$ & $2.09\pm 0.12$ & $2.26\pm 0.08$ & $3.16\pm 0.15$\\
    &ACC            & $2.75\pm 0.24$ & $1.92\pm 0.12$ & $ {2.48\pm0.25}$ & $ {3.78\pm0.50}$ & $3.14\pm 0.13$\\
    &WZ         & $2.52\pm 0.22$  & $1.73\pm 0.13$ & $2.20\pm 0.18$ & $2.26\pm 0.12$ & $3.31\pm 0.26$\\ 
    \toprule
        &        & 4 PM  & 5 PM & 6 PM & 7 PM & 8 PM\\
    \midrule
      \textbf{ \multirow{3}{*}{\rotatebox[origin=c]{90}{\tiny Forecast}}} &Normal   &  $5.10\pm 0.40$  & $4.09\pm 0.44$ & $3.16\pm 0.23$ & $2.01\pm 0.12$ & $1.54\pm 0.10$\\
    &ACC            & $5.15\pm 0.26$  & $3.94\pm0.41$ & $2.95\pm0.15$ & $1.95\pm0.12$ & $1.40\pm 0.06$\\
    &WZ           & $4.87\pm 0.05$  & $3.97\pm 0.17$ & $3.37\pm 0.15$ & $1.98\pm 0.07$ & $1.60\pm 0.17$ \\
    \midrule
       \textbf{ \multirow{3}{*}{\rotatebox[origin=c]{90}{\tiny Reconstruct}}}   &Normal& $4.53\pm 0.19$  & $3.69\pm 0.19$ & $2.60\pm 0.07$ & $1.89\pm 0.19$ & $1.80\pm 0.12$\\
    &ACC           & $4.60\pm 0.41$ & $3.62\pm 0.24$ & $2.74\pm 0.22$ & $1.99\pm 0.16$&$1.67\pm 0.09$ \\
    &WZ         & $4.91\pm 0.18$  & $3.69\pm 0.48$ & $2.70\pm 0.17$ & $2.01\pm 0.16$ & $1.72\pm 0.15$\\ 
    \toprule
        &        & 9 PM  & 10 PM & 11 PM &  & \\
    \midrule
      \textbf{ \multirow{3}{*}{\rotatebox[origin=c]{90}{\tiny Forecast}}} &Normal   & $1.43\pm 0.09$  & $1.20\pm 0.10$ & $1.01\pm 0.09$ &  & \\
    &ACC           &$1.50\pm 0.16$  & $0.95\pm 0.06$ &$0.85\pm 0.10$ &  &  \\
    &WZ           & $1.46\pm 0.13$  & $1.05\pm 0.08$ & $1.03\pm 0.06$ &  & \\
    \midrule
       \textbf{ \multirow{3}{*}{\rotatebox[origin=c]{90}{\tiny Reconstruct}}}   &Normal& $1.36\pm 0.11$  & $1.32\pm 0.09$ & $1.06\pm 0.11$ &  & \\
    &ACC          & $1.37\pm 0.07$ & $1.40\pm 0.15$ &$0.95\pm 0.10$  &  &\\
    &WZ         &$1.44\pm 0.09$ & $1.26\pm 0.18$ & $1.09\pm 0.04$&  & \\ 
    \bottomrule
    \end{tabular}
    \caption{The MAPE results of forecasting and reconstruction of SSMs per hour. The MAPE results are presented in three cases: Normal, ACC, and WZ. ACC is located on Edge 26-22, between 1:15 PM and 2:15 PM. WZ is located on Edge 22-21, between 6:00 AM and 8:00 AM. Despite these various cases and time-varying traffic patterns, the $ForecastingMAPE$ maintains the range between 0.85\% and 12.97\%, the $ReconstructionMAPE$ maintains the range between 0.95\% and 8.27\%.}
    \label{tab:ssm-mape}
\end{table}

\Cref{tab:ssm-mape} presents the MAPE results of SSMs across three scenarios: Normal, ACC, and WZ. The $ForecastingMAPE$ maintains the range between $0.85\%$ and $12.97\%$, the $ReconstructionMAPE$ maintains the range between $0.95\%$ and $8.27\%$. Both $ForecastingMAPE$ and $ReconstructionMAPE$ show similar values throughout the day, except during WZ and ACC occurrences. During these incidents, MAPE values were notably higher. Specifically, during the WZ incident, which closed two out of three lanes on Edge 22-21 between 6 AM and 8 AM, the MAPE values for both forecasting and reconstruction were elevated, compared to the Normal scenario, due to increased safety risks (higher SSMs). The MAPEs during WZ were approximately $0.5\%-2\%$ higher than the Normal scenario. Even at 9 AM, an hour after the WZ was cleared, the MAPEs remained slightly higher than Normal until 10 AM. This indicates that the post-impact of WZ continued to affect drivers, showing that traffic did not return to normal until 10 AM. Similarly, both $ForecastingMAPE$ and $ReconstructionMAPE$ were slightly higher than Normal and WZ during the ACC at 1 PM ($2.68\%$ and $2.48\%$) and 2 PM ($5.10\%$ and $3.78\%$), indicating that the ACC caused increased SSMs, leading to higher MAPE values.

\begin{figure}
    \centering
    \includegraphics[width=1\textwidth]{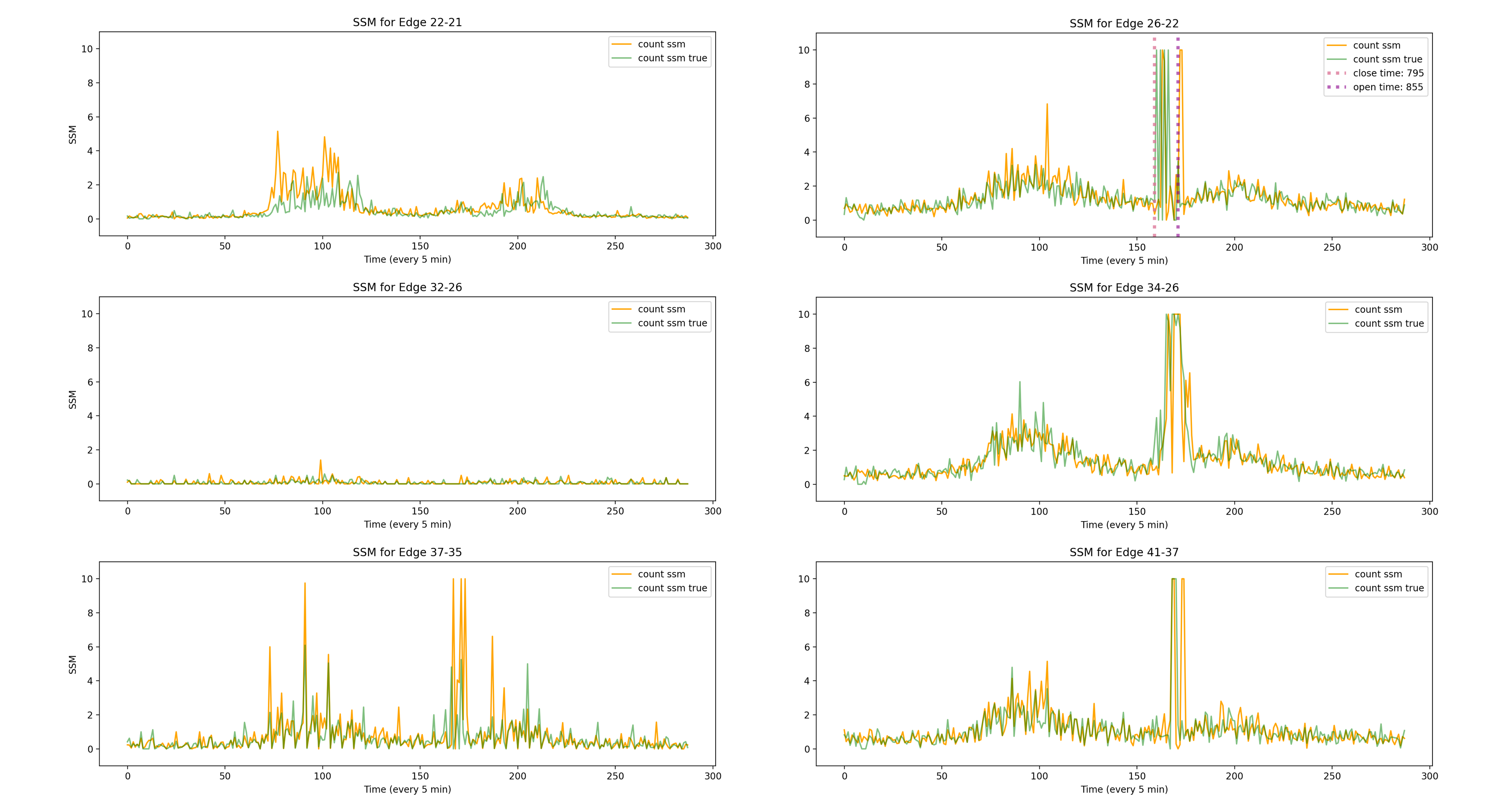}
    \caption{Real-time SSM prediction under ACC scenario. Green line represents true SSM, and yellow line represents the predicted SSMs by MTSS. The ACC occurred on Edge 26-22 from 1:15 PM to 2:15 PM, with a two-lane closure out of three lanes.}
    \label{fig:ssm_pre_mtss}
\end{figure}
\begin{figure}[!h]
\begin{subfigure}{1\textwidth}
    \centering
    \includegraphics[width=1\textwidth]{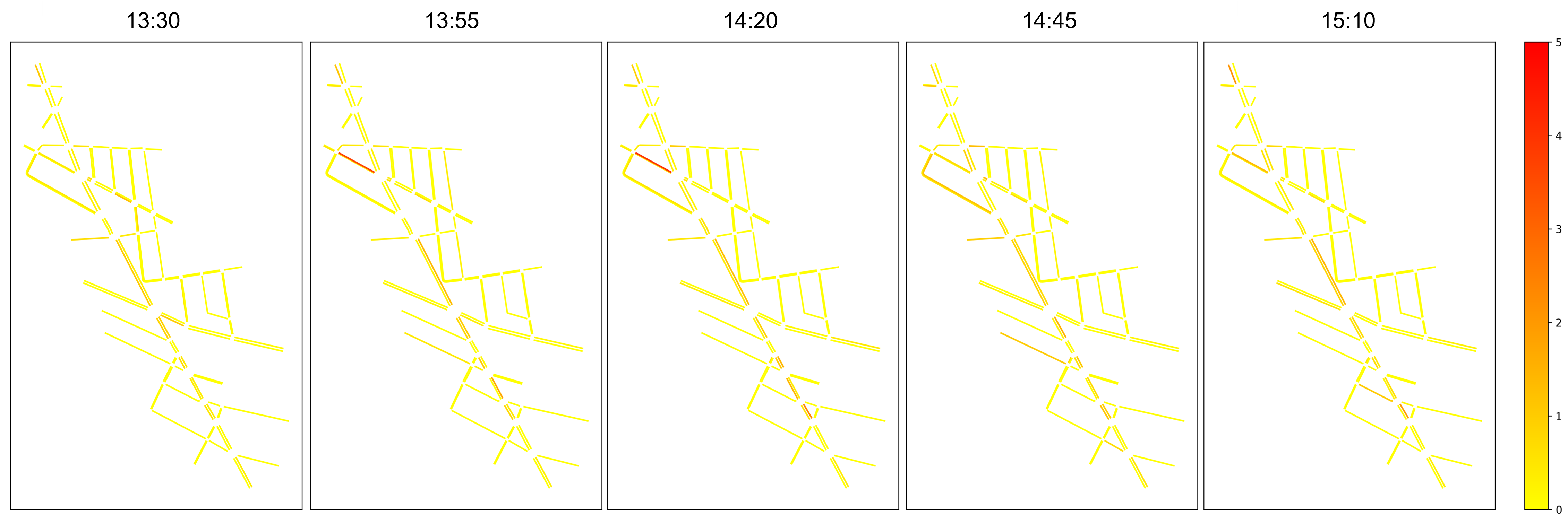}
    \caption{Normal}
\end{subfigure}
\begin{subfigure}{1\textwidth}
    \centering
    \includegraphics[width=1\textwidth]{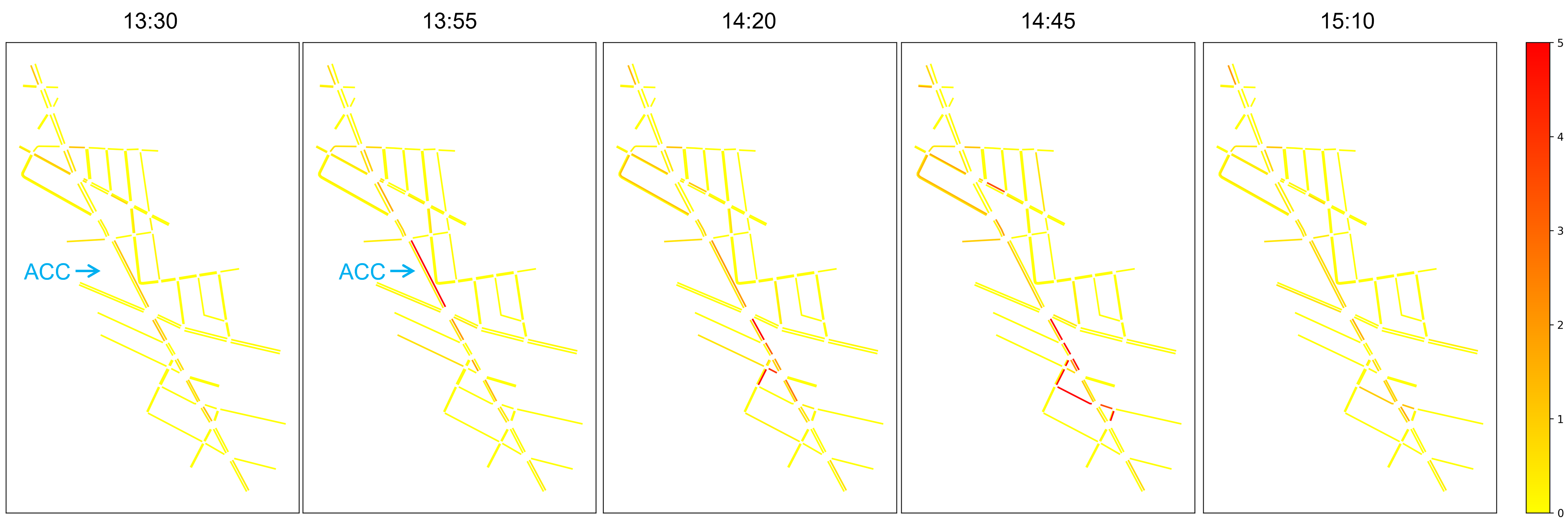}
    \caption{ACC}
\end{subfigure}
\caption{Heatmaps of predicted safety risks in comparison with Normal and ACC cases. The selected heatmaps cover timestamps from 1:30 PM to 3:10 PM. The upper row shows (from left to right) the propagation of predicted safety risks in normal conditions, the lower row shows (from left to right) the propagation of predicted safety risks during the occurrence of ACC.}
\label{fig:ssm_heatmap_acc}
\end{figure}
Further insights can be drawn from \Cref{fig:ssm_pre_mtss} and \Cref{fig:ssm_heatmap_acc}. \Cref{fig:ssm_pre_mtss} highlights selected edges where the ACC occurred and its neighboring edges. The yellow line represents predicted SSMs by the MTSS component, and the green line represents true SSMs. The predicted SSMs align well with the true SSMs, indicating good performance of the MTSS component in predicting safety risks under unexpected ACC situations. \Cref{fig:ssm_heatmap_acc} illustrates the propagation of SSMs during and after the ACC. Compared to the Normal scenario, the ACC caused increased SSMs on Edge 26-22 and its neighboring edges, with the increase spreading upstream along the main corridor (Flatbush Ave.) and minor roads. Even after the ACC was cleared at 14:15 PM (time step 170 in \Cref{fig:ssm_heatmap_acc}), elevated SSMs persisted and expanded to a larger area, indicating the post-impact of the ACC in both spatial and temporal dimensions.

The spatial and temporal evolution of predicted SSMs under traffic incident scenarios (e.g., ACC and WZ) provides transportation managers with valuable foresight. This enables them to anticipate potential high safety risks, identify road segments prone to hazard build-up, and allocate resources proactively to manage safety risks effectively.

\subsubsection{Evaluation of Risk-aware Multi-PTC Control}
\begin{figure}
    \centering
    \includegraphics[width=1\textwidth]{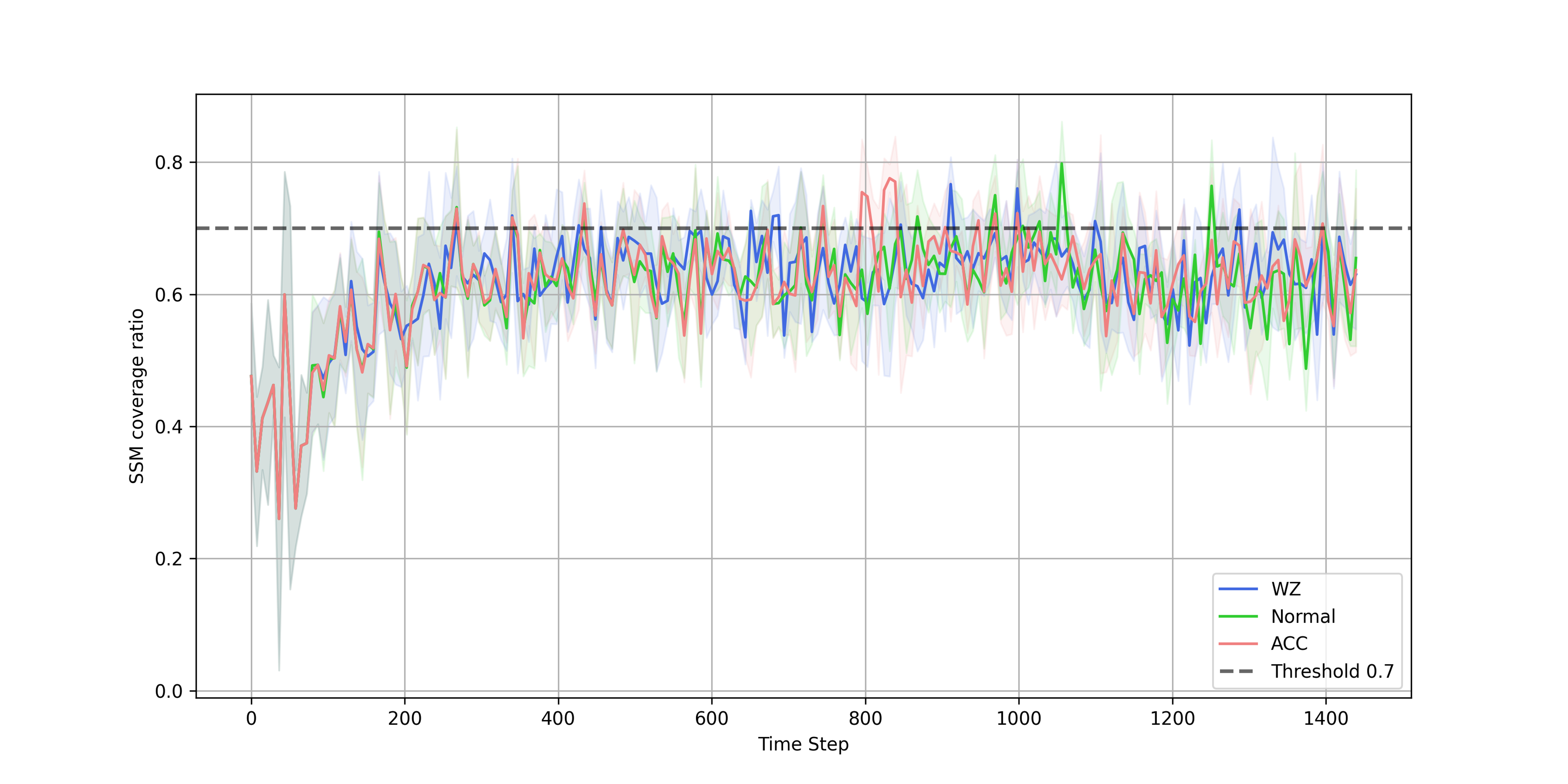}
    \caption{SSM coverage ratios in three scenarios. The dashed line indicates the 0.7 threshold in the constraint. }
    \label{fig:ssm-ratio}
\end{figure}

The objective of online PTC control, as delineated in \Cref{subsec:camera-surveillance}, is to monitor the edges with mobility and safety risks. To evaluate the RiCCOL's fulfillment of this objective, we begin with the real-time safety-risk coverage ratio, i.e., the percentage of actual SSM (i.e., $\bm{r}_t$) on the captured edges over the network's total SSM, which is given by $g(\bm{a}_t;\bm{r}_t)\triangleq \langle \bm{a}_t, \bm{r}_t \rangle/\|\bm{r}_t\|$ appearing as a constraint in \Cref{eq:constrained-opt}. \Cref{fig:ssm-ratio} summarizes the coverage ratio across three scenarios, where all ratios stay between $0.6$ and $0.7$ for most of the day. Even though the constraint in \Cref{eq:constrained-opt} requires more than $70\%$ percent coverage, we regard the results in \Cref{fig:ssm-ratio} satisfactory for the following reasons. First, the actual safety risk is unobservable to the PTCs, and RiCCOL uses simulated SSM as a surrogate, which inevitably brings in approximation error as presented in \Cref{tab:ssm-mape}. Such approximation errors may mislead the primal-dual update in RiCCOL, degrading the safety-risk coverage. Despite the loss of fidelity caused by the SSM, RiCCOL can still maintain a high ratio for most of the day.   
Second, RiCCOL applies online gradient descent and ascent to the primal and dual variables, respectively. Since the Lagrangian transforms the hard constraint into the soft one through the multiplier, each gradient update aims to increase the objective value and reduce the constraint violation without enforcing hard constraints on the action selection. \Cref{fig:ssm-ratio} shows that the ratios quickly rise above $0.6$ in around 200 minutes and stay within the range from $0.6$ to $0.7$ thereafter, indicating that the multi-PTC, powered by RiCCOL, is able to adapt to the dynamic traffic situations quickly.   

In addition to safety-risk-oriented surveillance, multi-PTC's timely responsiveness to various traffic incidents is also crucial in achieving real-time synergy between the physical world and the digital twin. To quantify the responsiveness, we introduce the notion \textit{time-to-sight} (TTS), referring to the time until the incident falls within the PTCs' sight, i.e., there exists at least one PTC monitoring the edge where the incident takes place. Given a traffic incident on the edge $(i,j)$, denoted by $Z_t(i,j)$, the  TTS is defined by the following
\begin{equation}
\label{eq:tts}
    \text{TTS}\triangleq\left\{\begin{aligned}
          & t-\min_{t'}\{t'\leq t| \bm{a}_\tau(i,j)=1, \forall \tau\in [t',t] \}, \quad \text{if }  \bm{a}_t(i,j)=1,\\
        & t- \min_{t'}\{t< t'| \bm{a}_{t'}(i,j)=1\},\quad \text{otherwise}.
    \end{aligned}\right.
\end{equation}
Of particular note is that TTS is a signed real value, where the positive value denotes the lead time (tilting PTCs preemptively before incidents happen), and the negative represents the lag time (tilting PTCs in response to incidents). A zero TTS indicates a fast response to traffic incidents without any lag, and the larger the TTS is, the more proactive the PTCs are. We report the TTS results under the ACC setup: the incidents are unexpected, which is an ideal testbed to evaluate the RiCCOL's real-time responsiveness under unanticipated traffic conditions. The TTS on the lane-closure and lane-reopening are $4.00\pm 5.76$ and $5.80\pm 5.19$ (minutes), respectively. It is straightforward to see that the positive TTS reveals RiCCOL's sensitivity to traffic disturbances, enabling timely monitoring of incidents. The preemptive PTC tilting is credited to the dual update in RiCCOL, which mandates the PTCs to cover edges with high safety risks before incidents occur.   

\subsubsection{Ablation experiment on LSTT}
The discussions above have focused on the MTSS's ability to predict and reconstruct the safety-risk information (see \Cref{tab:ssm-mape}), where LSTT plays an important part in adjusting the simulation period to evolving traffic conditions. This subsection aims to inspect to what extent the MTSS's effective operation depends on LSTT. Since LSTT directly affects MTSS and its simulated SSMs, our ablation study compares the MAPEs of SSM predictions returned by MTSS under LSTT and a 60-minute (long-term) fixed simulation period. In all repeated experiments, the short-term twining is triggered three times in a day: from 6 AM to 7 AM, from 1:30 PM to 4 PM, and from 5 PM to 7 PM. The first and last short-term twinings are caused by the increased demand during rush hours, while the second one is due to the accident lasting from 1:15 PM to 2:15 PM.  

\begin{figure}[!h]
\begin{subfigure}{1\textwidth}
    \centering
    \includegraphics[width=1\textwidth]{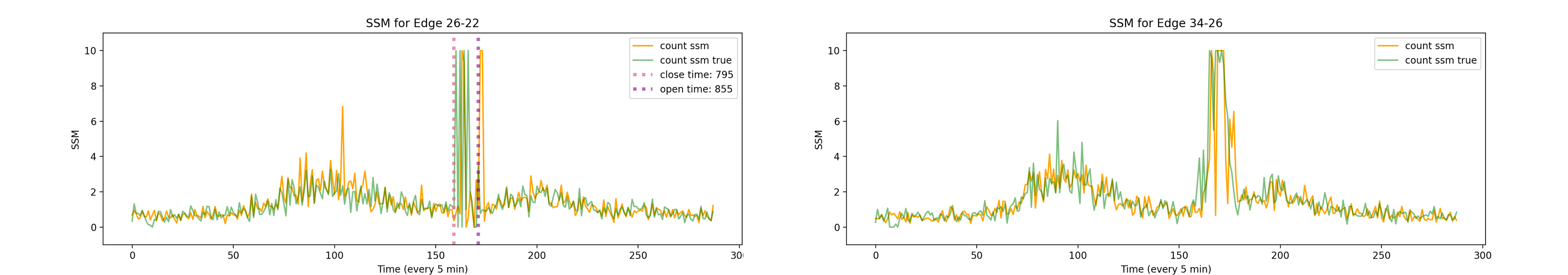}
    \caption{Real-time SSM prediction under LSTT}
    \label{fig:ssm-plot-acc-lstt}
\end{subfigure}
\begin{subfigure}{1\textwidth}
    \centering
    \includegraphics[width=1\textwidth]{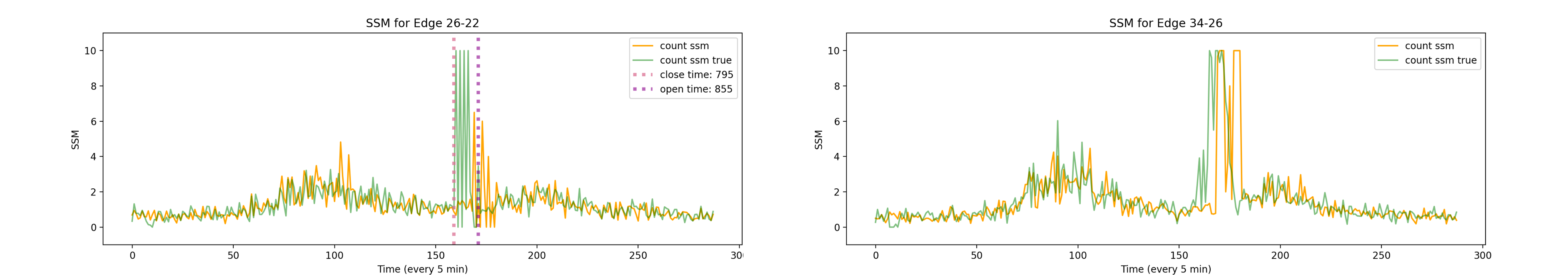}
    \caption{Real-time SSM prediction under Long-term twining}
    \label{fig:ssm-plot-acc-long}
\end{subfigure}
\caption{A comparison of real-time SSM prediction under LSTT and long-term twining. Since long-term twining fails to update the accident information timely, the corresponding SSM prediction lags behind the actual SSM evolution, when the accident takes place at 1:15 PM (160th step in the x-axis).}
\label{fig:ssm-plot-acc}
\end{figure}

\Cref{tab:ablation-lstt-ssm} presents the comparison of SSM prediction MAPEs under two twining patterns. One can see that when short-term twining is in operation, the resulting MAPEs are smaller than those under long-term twining (see the bold numbers). In particular, the SSM prediction under long-term twining is worse than that under LSTT by a large margin (around $2\%$) at 2 PM. This is because long-term twining is unable to update the lane-reopening information in the digital twin on time: it is until 3 PM that the MTSS receives the reopening information. \Cref{fig:ssm-plot-acc} gives a more direct visualization of the misalignment between the DT and the real world under the long-term twining. One can see that there is a shift between the peaks of two colored lines after the accident takes place at 1:15 PM (the 160th step in the x-axis), indicating that long-term twining fails to update the accident information promptly and the resulting SSM predictions deviate from the reality significantly. In contrast, LSTT provides more accurate SSM predictions, as shown in \Cref{fig:ssm-plot-acc-lstt} by activating the short-term twining when the accident is in effect. Therefore, LSTT offers a self-adaptive mechanism that enables MTSS to produce accurate replicas with minimal latency, creating a real-time synergy between the physical world and the digital twin while maintaining low-cost operation.     

\begin{table}[!h]
    \centering
    \begin{tabular}{cccccc}
     \toprule
         MAPE ($\%$) & 1 AM & 2 AM & 3 AM & 4 AM & 5 AM \\
        \midrule
         Long &  $1.02 \pm 0.11$ & $1.14 \pm 0.21$ & $1.03 \pm 0.13$ & $1.43 \pm 0.22$ & $2.19 \pm 0.10$\\
         LSTT & $1.01\pm0.11$ & $1.10\pm 0.18$ & $1.02\pm 0.14$ & $1.37\pm 0.15$& $2.23\pm 0.12$ \\
         \toprule
         & 6 AM & 7 AM & 8 AM & 9 AM & 10 AM \\
         \midrule
    Long & $6.28 \pm 0.48$ & $9.13\pm0.56$ & $8.97 \pm 0.58$ & $4.88 \pm 0.11$ &  $3.02 \pm 0.10$\\
    LSTT  & $\mathbf{6.23\pm 0.37}$  &$\bm{8.58 \pm 0.70}$  & $8.92\pm0.52$ & $5.02\pm 0.30$ & $2.74\pm 0.14$\\
    \toprule
        & 11 AM & 12 PM & 1 PM  & 2 PM & 3 PM\\
        \midrule
    Long & $2.76 \pm 0.23$ & $2.14 \pm 0.10$ & $3.65 \pm 0.22$ & $7.57 \pm 0.28$ & $4.82 \pm 0.57$\\
    LSTT & $2.57\pm 0.15$  & $1.95\pm0.15$ & $\bm{2.68\pm 0.17}$ &$\bm{5.10\pm0.11}$ & $\bm{3.86\pm 0.32}$\\
    \toprule
        & 4 PM  & 5 PM & 6 PM & 7 PM & 8 PM\\
        \midrule
    Long& $5.49 \pm 0.23$ & $4.01 \pm 0.24$ & $2.97 \pm 0.13$ & $1.88 \pm 0.08$ &  $1.52 \pm 0.08$\\
    LSTT  & $\bm{5.15\pm 0.26}$  & $\bm{3.94 \pm 0.41}$ & $\bm{2.95\pm0.15}$ & $1.95\pm0.12$ & $1.40\pm 0.06$\\
    \toprule
        & 9 PM  & 10 PM & 11 PM &  & \\
        \midrule
    Long & $1.54 \pm 0.06$ & $1.07 \pm 0.04$ & $0.90 \pm 0.05$ & &\\
    LSTT &$1.50\pm 0.16$  & $0.95\pm 0.06$ &$0.85\pm 0.10$ &  &  \\
    \bottomrule
    \end{tabular}
    \caption{Real-time SSM prediction MAPEs under long-term twining and LSTT. In LSTT, the short-term twining is activated from 6 AM to 7 AM, from 1:30 PM to 4:00 PM, and from 5:00 PM to 7:00 PM. The corresponding MAPEs by LSTT are bolded. The accident lasts from 1:15 PM to 2:15 PM.  LSTT returns smaller MAPEs for most of the day and, particularly, improves the SSM prediction by $2\%$ at 2:00 PM when the accident is in effect.}
    \label{tab:ablation-lstt-ssm}
\end{table}

\subsection{DT-DIMA use case: Hassle-free D.M. Scenarios}
One direct application of DT-DIMA is to provide hassle-free decision-making scenarios for the real-time management of urban mobility systems. By replicating the physical world and predicting future traffic situations, the DT-DIMA system equips transportation managers with proactive insights for their management strategies, identifying corresponding safety risks.

In this section, we illustrate a use case where transportation managers use the DT-DIMA system to manage a work zone in real time. Specifically, we modified the WZ case described in \Cref{subsec:experiment_setup} into three scenarios:

\begin{itemize}
    \item \textbf{WZ 2-0}: Work zone on Edge 22-21 closes two lanes from 6 AM to 7 AM, then reopens all lanes from 7 AM.
    \item \textbf{WZ 2-1}: Work zone on Edge 22-21 closes two lanes from 6 AM to 7 AM, then closes one lane from 7 AM to 8 AM.
    \item \textbf{WZ 2-2}: Original work zone, closing two lanes from 6 AM to 8 AM.
\end{itemize}

\begin{figure}[htp]
\centering
\includegraphics[width=0.95\textwidth]{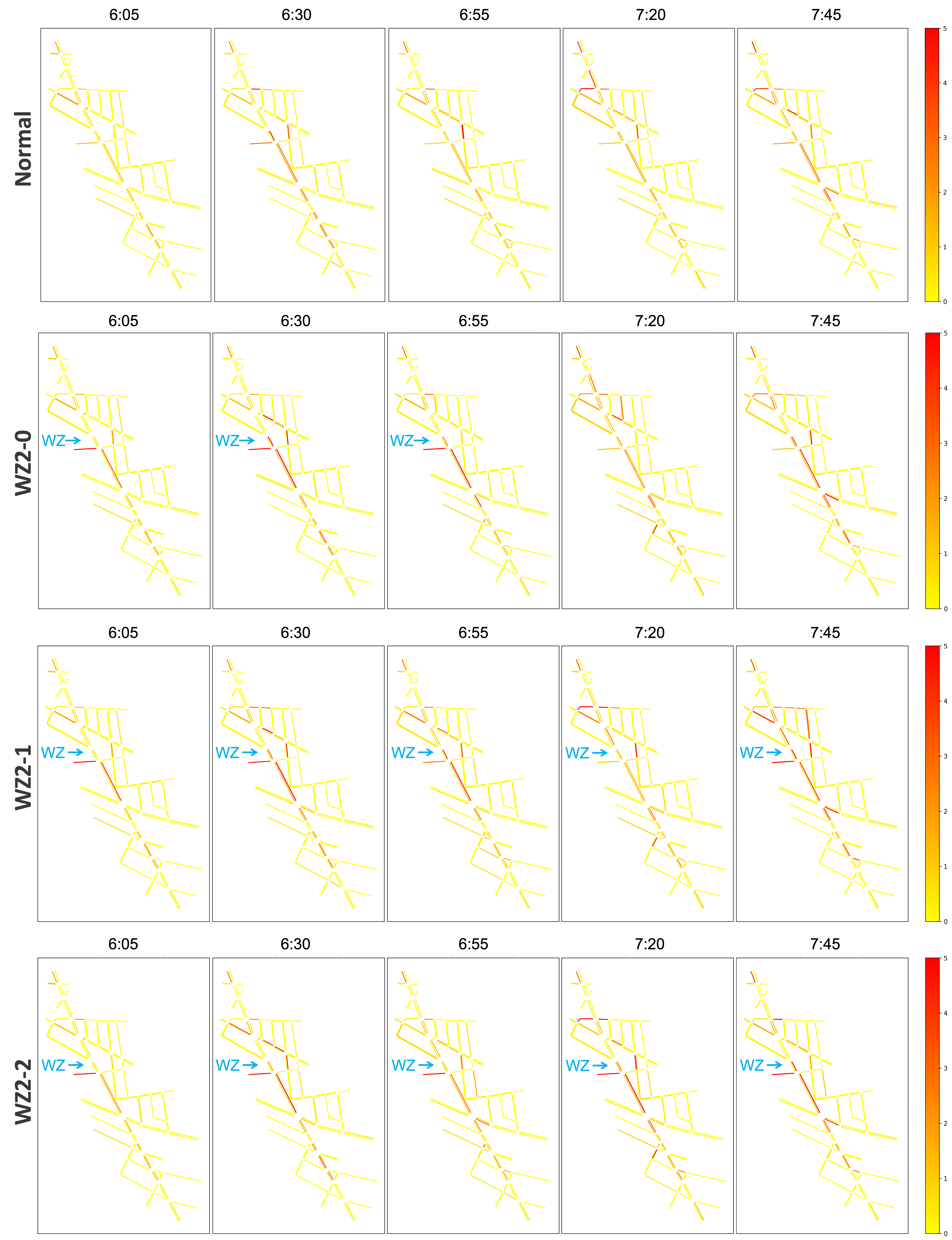}
\caption{Heatmaps of predicted safety risks in comparison with Normal, WZ 2-0, WZ 2-1, WZ 2-2 cases. The timestamp index uses a 5-min aggregation level, time steps 73, 78, 83, 88, and 93 present at 6:08 AM, 6:33 AM, 6:58 AM, 7:23 AM, 7:48 AM, respectively.}
\label{fig: ssm-heatmap-hasslefree}
\end{figure}

Using these scenarios, we demonstrate how the DT-DIMA system allows transportation managers to run multiple scenarios seamlessly and without external costs during an ongoing work zone. After the initial one-hour two-lane closure, managers can revisit the original work zone configurations to assess if changes are needed to minimize traffic disruption. As shown in \Cref{fig: ssm-heatmap-hasslefree}, DT-DIMA can simultaneously and seamlessly run WZ 2-0, WZ 2-1, and WZ 2-2 scenarios, providing managers with the propagation of safety risks in both spatial and temporal dimensions for each configuration.

It was found that if managers choose to continue lane closures (either one or two lanes) during the second hour (7 AM to 8 AM), safety risks will propagate and expand from the main Flatbush corridor to the surrounding neighborhoods. Conversely, if the work zone is removed and all lanes are reopened during the second hour, safety risks along the Flatbush corridor will persist, but the surrounding neighborhoods will not be as significantly affected as in the WZ 2-1 and WZ 2-2 scenarios.

This use case underscores the power of the DT-DIMA system in assisting transportation managers with managing mobility systems under both normal and incidental situations. It allows them to foresee the evolution and propagation of safety risks, enabling proactive resource allocation to mitigate the negative impacts of their management strategies on drivers.

\section{Implication and future work}
The current DT-DIMA system relies on limited PTCs at key intersections as traffic sensing sources. An important implication of the DT-DIMA system is its potential to integrate with other urban sensing infrastructures, such as loop detectors, WiFi/Bluetooth sensors, and connected automated vehicles (CAVs). For example, incorporating traffic information from loop detectors would enable DT-DIMA to gather more comprehensive traffic flow data along road segments, enhancing the accuracy of traffic information sent to the DT. This, in turn, allows RiCCOL to focus more on monitoring high safety risks. Additionally, DT-DIMA can collaborate with CAVs to obtain real-time, driver-centric information, reducing the reliance solely on the MTSS component for safety risk data. Despite their low penetration rate, CAVs can still work cooperatively with DT-DIMA to identify road hazards and other safety-related concerns experienced by drivers.

Currently, DT-DIMA intervenes in the physical world primarily by controlling PTCs at intersections. A potential enhancement is to incorporate more traffic control infrastructures, such as signal control, variable speed limits, and dynamic message signs. For example, by linking signal control with predicted safety risks, DT-DIMA can adaptively optimize signal phases to reduce drivers' risk exposure to unexpected accidents. Furthermore, DT-DIMA can integrate with other traffic control mechanisms to detour and reroute traffic, helping to avoid areas of high safety risk.

\section{Conclusion}
In this paper, we proposed a digital twin-based system for urban mobility management named DT-DIMA. By leveraging the existing PTC infrastructure at intersections within the road network, DT-DIMA acquires real-time traffic information in a cost-efficient manner. This information is then integrated into a driver-informed predictive service. The STTE component of this service fuses observed data from PTCs to generate network-wide traffic information, predicting future traffic flows based on this fused data. The MTSS component uses this network-wide traffic information to replicate the physical world and predict safety risks along each road segment in the network. Additionally, we designed a novel distributed and cooperative PTC control algorithm that optimizes PTC control by considering both the predicted traffic flow from STTE and the safety risks from MTSS. This ensures effective monitoring of road segments with high safety and mobility risks. We compare our results with a base model, which uses the current tilting strategy of PTCs in the real world. Our experiments across three case scenarios—normal conditions, an unexpected accident, and a pre-planned work zone—demonstrated the efficacy of the DT-DIMA system. As compared with the base model, the STTE component showed superior performance by achieving mean absolute percentage errors (MAPEs) in estimating the current traffic states ranging from $8.40\%$ to $15.11\%$ as compared with $40.63\%$ to $43.94\%$ from the base model. The MTSS component achieved mean absolute percentage errors (MAPE) ranging from $0.85\%$ to $12.97\%$ in forecasting network-level safety risks. During pre-planned work zones and unexpected accident scenarios, the MTSS component showed a $0.5-2\%$ increase in MAPE as compared with normal conditions. Powered by the driver-informed predictive service, the proposed RiCCOL control preemptively tilts PTCs to monitor edges with high driving risks before incidents occur. Experiments demonstrate that RiCCOL achieves an intra-PTC coordination that effectively monitors around $70\%$ of hazardous roads and creates a $5$-minute lead time in capturing traffic incidents. Finally, we showcased a use case demonstrating DT-DIMA's potential in managing mobility systems in real-time under incidental situations. DT-DIMA can forecast the propagation of safety risks in both spatial and temporal dimensions, responding to different management strategies. This allows transportation managers to anticipate the evolution and propagation of safety risks, enabling proactive resource allocation to mitigate the negative impacts experienced by drivers due to their management strategies.

\section{Acknowledgements}
This work was supported by the C2SMARTER, a Tier 1 U.S. Department of Transportation (USDOT) funded University Transportation Center (UTC) led by New York University funded by USDOT. The contents of this paper only reflect the views of the authors who are responsible for the facts and do not represent any official views of any sponsoring organizations or agencies.

\section{Author Contributions}
The authors confirm their contribution to the paper as follows. Conceptualization: Zilin Bian, Tao Li, Fan Zuo, Kaan Ozbay; Data curation: Fan Zuo and Haozhe Lei; Formal analysis: Tao Li, Zilin Bian, Haozhe Lei; Methodology: Tao Li, Zilin Bian, Haozhe Lei, Fan Zuo, Ya-Ting Yang, Kaan Ozbay; Validation: Tao Li, Zilin Bian, Haozhe Lei; Visualization: Ya-Ting Yang, Fan Zuo, Tao Li, Zilin Bian; Writing - original draft: Zilin Bian, Tao Li, Fan Zuo, Ya-Ting Yang; and Writing - review \& editing: Zilin Bian, Tao Li, Fan Zuo, Kaan Ozbay, Quanyan Zhu, Zhenning Li, Zhibin Chen. All authors reviewed the results and approved the final version of the manuscript.

\bibliography{ref} 
\bibliographystyle{unsrt}
\end{document}